\documentclass[aps,prl,superscriptaddress,twocolumn,showpacs]{revtex4-1}

\usepackage[utf8]{inputenc}
\usepackage[T1]{fontenc}
\usepackage[german,english]{babel}
\usepackage{amsmath}
\usepackage{mathtools}
\usepackage{amssymb}
\usepackage{amsthm}
\usepackage{tensor}
\usepackage{microtype} 
\usepackage{quoting} 
\quotingsetup{font=small}

\usepackage{graphicx}
\usepackage{fancyhdr}
\usepackage{braket}
\usepackage{bm}
\usepackage{comment}
\usepackage{booktabs}
\usepackage{eurosym} 
\usepackage{braket}
\usepackage{bbold} 
\usepackage{emptypage}
\usepackage{epstopdf}
\usepackage{verbatim}
\usepackage{enumitem}
\usepackage{float} 
\usepackage{chemformula}

\definecolor{islamicgreen}{rgb}{0.0, 0.56, 0.0}

\newcommand{\beq}{\begin{equation}}
\newcommand{\eeq}{\end{equation}}

\usepackage[bookmarks=true,colorlinks,linkcolor=blue,urlcolor=blue,citecolor=blue]{hyperref} 

\begin{document}

\title{
Designing non-equilibrium states of quantum matter through stochastic resetting
}
\author{Gabriele Perfetto}
\affiliation{Institut f\"ur Theoretische Physik, Eberhard Karls Universit\"at T\"ubingen, Auf der
Morgenstelle 14, 72076 T\"ubingen, Germany.}
\author{Federico Carollo}
\affiliation{Institut f\"ur Theoretische Physik, Eberhard Karls Universit\"at T\"ubingen, Auf der
Morgenstelle 14, 72076 T\"ubingen, Germany.}
\author{Matteo Magoni}
\affiliation{Institut f\"ur Theoretische Physik, Eberhard Karls Universit\"at T\"ubingen, Auf der
Morgenstelle 14, 72076 T\"ubingen, Germany.}
\author{Igor Lesanovsky}
\affiliation{Institut f\"ur Theoretische Physik, Eberhard Karls Universit\"at T\"ubingen, Auf der
Morgenstelle 14, 72076 T\"ubingen, Germany.}
\affiliation{School of Physics and Astronomy and Centre for the Mathematics and Theoretical
Physics of Quantum Non-Equilibrium Systems, The University of Nottingham, Nottingham,
NG7 2RD, United Kingdom.}

\begin{abstract}
We consider closed quantum many-body systems subject to stochastic resetting. This means that their unitary time evolution is interrupted by resets at randomly selected times.
When a reset takes place the system is reinitialized to a state chosen from a set of reset states conditionally on the outcome of a measurement taken immediately before resetting. We construct analytically the resulting non-equilibrium stationary state, thereby establishing an explicit connection between quantum quenches in closed systems and the emergent open system dynamics induced by stochastic resetting. We discuss as an application the paradigmatic transverse-field quantum Ising chain. We show that signatures of its ground-state quantum phase transition are visible in the steady state of the reset dynamics as a sharp crossover. Our findings show that a controlled stochastic resetting dynamics allows to design non-equilibrium stationary states of quantum many-body systems, where uncontrolled dissipation and heating can be prevented. These states can thus be created on demand and exploited, e.g., as a resource for quantum enhanced sensing on quantum simulator platforms. 
\end{abstract}

\maketitle

\emph{Introduction.} Deterministic or stochastic dynamics interspersed with stochastic resetting leads to the emergence of a non-equilibrium steady state (NESS). This concept has been put forward in the classical realm in Refs.~\cite{evans2011resetting,evans2011resettinglong,evans2014diffusion,DPTreset2015,eule2016non}, where it has been established that a particle undergoing Brownian motion reaches a NESS when its diffusive time evolution is \textit{reset} at constant rate $\gamma$ (Poissonian reset) to its initial position. 
Using the renewal equation approach \cite{evans2011resetting,evans2011resettinglong,renewal1,renewal2,renewal3,renewal4,renewal5,renewal6,renewal7,magoni2020}, it has been further shown that stochastic resetting can improve the efficiency of search processes in terms of their mean first passage time \cite{evans2011resetting,evans2011resettinglong,redner2001guide,bray2013persistence,evans2013optimal,Montero2013,christou2015diffusion,Campos2015,montero2017continuous,Maso2019,Tucci2019,evans2020review,busiello2021}. 
In contrast to the case of classical systems, much less is known about the effect of stochastic resetting on quantum systems, either closed or open. In the latter case, the Markovian Lindblad equation  \cite{lindblad1976generators,gorini1976completely} describing the evolution in the presence of Poissonian resetting has been analyzed in Refs.~\cite{Hartmann2006,Linden2010,Rose2018spectral,carollo2019unravelling,Armin2020,riera2020measurement}. Closed quantum systems which are reinitialized to their initial state after Poissonian resetting 
have been, instead, analyzed without reference to the Lindblad equation in Ref.~\cite{Mukherjee2018}. Using the renewal equation approach, it was shown in the latter work that the reached NESS differs from the diagonal ensemble for any finite, non-zero, value of $\gamma$. For $\gamma \to \infty$, instead, one recovers the quantum Zeno effect \cite{zeno1977one,zeno1977two}, where the system does not evolve in time due to frequent reset projections. Resetting in closed quantum systems has been also studied in Refs.~\cite{Aharonov1993,meyer1996quantum,barkai2016,Baraki2017,Barkai2018,Barkai2020} for first detection problems. These recent works suggest that there is a direct link between the dynamics of closed quantum systems subject to stochastic resetting and an effective open system dynamics. However, the precise nature of this relation, and of possible connections between the NESS emerging from the resetting, unitary dynamics, and non-equilibrium signatures of quantum phase transitions  \cite{QPT1997,QPT2003,QPT2007} is not fully understood.

In this work we establish such a connection by developing a general theory based on semi-Markov processes \cite{janssen2006applied,carollo2019unravelling}. We prove analytically that by averaging the microscopic unitary time evolution over the reset distribution an effective non-Markovian open dynamics, governed by a generalized Lindblad equation, emerges. The ensuing NESS can be constructed analytically and its structure establishes a transparent connection to quantum quenches in closed systems, which can display signatures of equilibrium quantum phase transitions
\cite{Sachdev2004quenches,quench1,quench2,calabrese2011quantum,calabrese2012quantum1,calabrese2012quantum2,Heyl1,Heylreview}. Our results for the NESS recover the quantum Zeno effect in the limit of infinite resetting rate. We illustrate our ideas using, as an example, the quantum Ising chain in a transverse field (TFIC). The ground-state quantum phase transition of the model is signalled in the NESS by a crossover, whose sharpness is controlled by the distribution of the reset time. Our analysis establishes stochastic resetting as a tool to generate complex stationary many-body quantum states
without detrimental processes, such as heating, that typically occur when coherent dynamics is competing with dissipative processes.

\emph{Reset protocol} The results we are going to develop in the following hold for a general many-body quantum system made of $N$ particles or spin degrees of freedom, but for concreteness we will base our discussion mostly on the paradigmatic TFIC with Hamiltonian
\begin{equation}
H= -J \sum_{n=1}^{N} (\sigma_n^x \sigma_{n+1}^x +h\sigma_n^z).
\label{eq:TFIC_Hamiltonian}
\end{equation}
Here $\sigma_n^{x,y,z}$ are the Pauli matrices acting on site $n$, $J>0$ is the ferromagnetic coupling constant and $h$ is the strength of a transverse magnetic field. This model exhibits a quantum phase transition in its ground state at the critical field strength $h=h_c=1$, which can be characterized by using the  magnetization density $m=\sum_{n=1}^{N}\sigma_n^x/N$ as order parameter. 

\begin{figure}[t]
    \centering
    \includegraphics[width=0.8\columnwidth]{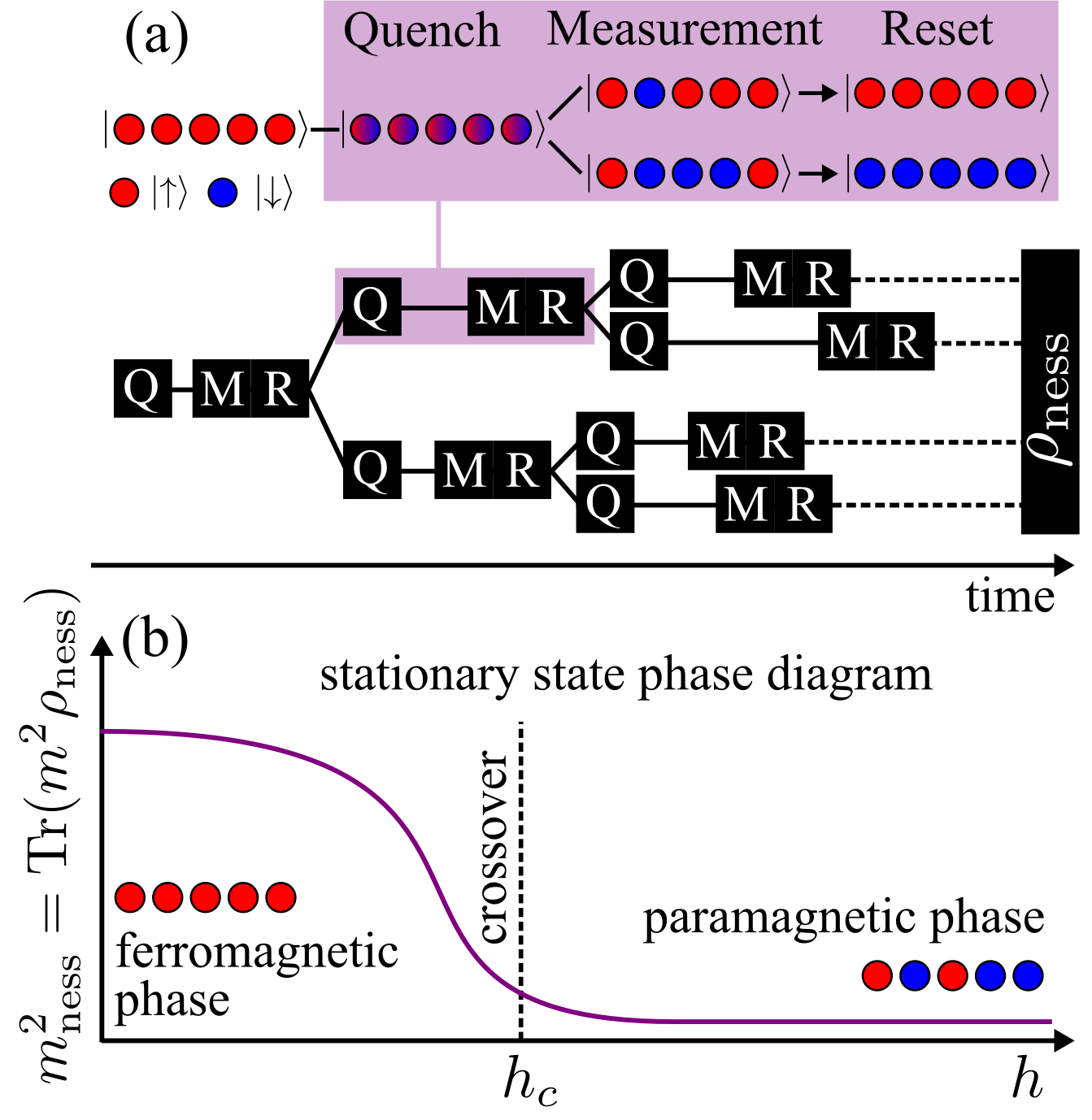}
    \caption{\textbf{Conditional reset protocol.} 
    (a) The time evolution is obtained upon alternating the quench-unitary dynamics, schematized by the ``Q'' boxes, and measurement-reset events, schematized by the ``M-R'' boxes, which happen at randomly selected times from a waiting time distribution. The average over all the ensuing ``reset trajectories'' generates the NESS density matrix $\rho_{\mathrm{ness}}$. In the inset (purple rectangle) the time evolution ensuing in a single quench dynamics and reset is shown. Conditioned on the measured value of the magnetization being positive or negative, the system is reinitialized either in a reset state completely polarized upwards ($\ket{1}=\ket{\uparrow \uparrow \dots \uparrow}$) or downwards ($\ket{2}=\ket{\downarrow \downarrow \dots \downarrow}$), respectively. (b) Sketch of the squared magnetization density $m^2_{\mathrm{ness}}$ in the NESS as a function of the transverse field $h$ [see Eqs.~\eqref{eq:TFIC_Hamiltonian} and \eqref{eq:squared_m_final}]. It displays remnants of the equilibrium quantum phase transition at ${h_c=1}$ through a sharp crossover between a ferromagnetic and a paramagnetic stationary phase.}
    \label{fig:1}
\end{figure}

We envisage a ``conditional reset'' protocol, which is sketched in Fig.~\ref{fig:1}(a): Starting from a given initial state, in the Figure the state $\ket{\psi_0}=\ket{\uparrow \uparrow \dots \uparrow}$ \footnote{The NESS does not depend on the specific choice of the initial state. The latter influences only the transient dynamics.}), the system is evolved unitarily under its many-body Hamiltonian $H$, e.g., Eq.~(\ref{eq:TFIC_Hamiltonian}) for the TFIC. After a randomly selected time, an observable is measured, in the Figure we measure the magnetization, and the many-body state is \textit{reset} to a state chosen within a set of ``reset states'' \textit{conditionally} on the outcome of the measurement. In our example the reset states are $\ket{1}=\ket{\uparrow \uparrow \dots \uparrow}$ and $\ket{2}=\ket{\downarrow \downarrow \dots \downarrow}$ (with $\ket{\uparrow}$ and $\ket{\downarrow}$ referring to the longitudinal-$x$ direction in Eq.~\eqref{eq:TFIC_Hamiltonian}). Our motivation to investigate the two aforementioned reset states is twofold: first, they are simple product states easy to prepare experimentally; second, they allow to establish a connection between the NESS of the reset dynamics and quantum quenches in the Ising chain. The two reset states can be indeed identified with the two degenerate ground states $\ket{\mbox{GS}(h=0)}_{\pm}$ of $H$ in Eq.~\eqref{eq:TFIC_Hamiltonian} for $h=0$, namely, $\ket{1}=\ket{\mbox{GS}(h=0)}_{+}$ and $\ket{2}=\ket{\mbox{GS}(h=0)}_{-}$. The unitary dynamics from the reset state $\ket{j}$ ($j=1,2$), $\ket{\psi_j(\tau)}=\mbox{exp}(-iH\tau)\ket{j}$ (setting $\hbar=1$), between consecutive resets stems therefore from a quench of the transverse field from the pre-quench $h_0=0$ to the post-quench value $h$ \cite{Sachdev2004quenches,quench1,quench2,quench4,quench5,calabrese2011quantum,calabrese2012quantum1,calabrese2012quantum2,quench6,Silva2008,Evangelisti2012,foini2012dynamic,quenchIsingExp,bucciantini2014quantum,Heyl1,Heylreview,refereeeIsing2021}. The experimental motivation behind the conditional reset protocol comes from recent results on Rydberg quantum simulators \cite{RydbergReview2020}, which allow to experimentally implement Ising models with transverse and longitudinal fields \cite{conf1,conf2,conf3,conf4,conf5,conf6,conf7,conf8} and permit, at the same time, the spatially resolved detection of the spin state upon which the reset is conditioned. Such measurement has been indeed implemented in Rydberg atom experiments on quantum quenches of the Ising model \cite{gasMicroscope2010,RydbergExpIsing1,RydbergExpIsing2,RydbergExpIsing3}. 

After the reset, the system evolves again under the quench dynamics up to the time at which the next measurement is performed. The time between consecutive measurements is drawn from a \textit{waiting time probability density}, $p(\tau)$, which is normalized as $\int_{0}^{\infty}\mbox{d}\tau \, p(\tau)=1$. Repeating this procedure leads at long times to a NESS, which in the case of the TFIC exhibits a sharp crossover between two stationary phases, as sketched in Fig.~\ref{fig:1}(b). Resetting can thus be interpreted as a dissipative process which leads to an effective open-system evolution. 

The reset protocol also allows to suppress naturally occurring dissipative processes, e.g. spontaneous photon emission or scattering in cold-atom experiments. This is achieved by choosing the probability $p(\tau)$ such that the reset must occur before a finite \textit{maximum reset time} $t_{\mathrm{max}}$, which is shorter than the dissipation timescale. This may allow to reduce decoherence and heating and thus enables one to tailor states of matter with robustness and long-term stability. In the following we will develop the theory for deriving the form of this NESS and we discuss specifically the case of the TFIC. Note, however, that although the present discussion is based on two reset states, our derivations can be generalized to the case of an arbitrary number of reset states (see Supplemental Material \cite{SM}).

\emph{Non-equilibrium stationary state} When a reset takes place, the system is reinitialized in the reset state $\ket{k}$ ($k=1,2$) dependently on the sign of the measured magnetization, see Fig. \ref{fig:1}(a). Which reset state is chosen depends only on the time $\tau$ elapsed since the previous reset event and on its outcome $\ket{j}$ ($j=1,2$), but, fundamentally, not on the previous history of the dynamics. The conditional reset dynamics can be therefore understood as a semi-Markov process \cite{janssen2006applied,carollo2019unravelling}, which is characterized by the transition probabilities $P_{jk}(\tau)$ (obeying $\sum_{k} P_{jk}(\tau)=1$) from the previous reset state $\ket{j}$ to the following reset state $\ket{k}$, given that a time $\tau$ has elapsed between the two resets. The formalism of semi-Markov
processes has been applied in Ref.~\cite{carollo2019unravelling} to Markovian open
quantum systems. Here we consider, instead, the case of an underlying unitary time evolution, and for the conditional reset dynamics sketched in Fig.~\ref{fig:1} we write
\begin{equation}
P_{jk}(\tau)=\Braket{\psi_j(\tau)| \mathcal{P}_k | \psi_j(\tau)},
\label{eq:P_ij_Projector}
\end{equation}
where $\mathcal{P}_k$ denotes the projector onto the set of states of the Hilbert space with positive ($k=1$) or negative ($k=2$) magnetization \footnote{For an even number $N$ of spins, $\mathcal{P}_k$ is not a projector since one has to account for states with zero total magnetization. This does not spoil our general theory since the fundamental property $\sum_k P_{jk}(\tau)=1$ still holds (see Supplemental Material \cite{SM}).}.

To describe the time evolution of the density matrix $\rho(t)$ we exploit the fact that the resetting protocol induces a \textit{renewal structure}, in the sense that the dynamics after a certain reset does not depend on what happened before that reset. This observation allows to derive ``renewal equations'' which express the dynamics of the state $\rho(t)$ in the presence of resets, in terms of the state $\rho_{\mathrm{free},j}(t)=\ket{\psi_j(t)}\bra{\psi_j(t)}$ which evolves under a dynamics where resets are absent. 
We stress that in our analysis the resetting is coupled to the underlying unitary dynamics via Eq.~\eqref{eq:P_ij_Projector}. This makes the conditional reset dynamics fundamentally different with respect to resets to a single state, first analyzed in Ref.~\cite{Mukherjee2018} for closed quantum systems, where the reset is independent from the underlying dynamics, which significantly simplifies the analysis.

We consider the observation time $t$ to be fixed and calculate the average density matrix $\rho_n(t)$ over all the ``reset trajectories'' that have exactly $n\geq 0$ resets within the interval $[0,t]$. The density matrix is then given by accounting for all the reset trajectories with an arbitrary number of resets $\rho(t)=\sum_{n=0}^{\infty} \rho_n(t)$, as shown pictorially in Fig.~\ref{fig:1}(a). The previous expression can be resummed by taking the Laplace transform $\widehat{\rho}(s)$ of $\rho(t)$. The average density matrix $\widehat{\rho}_n(s)$ is, indeed, obtained starting from the initial state, similarly as in Markov chains, by considering $n$ reset transitions $\widehat{R}^n(s)$, with the transition matrix $\widehat{R}_{jk}(s)$:  
\begin{equation}
\widehat{R}_{jk}(s) =\int_{0}^{\infty} \mbox{d}\tau \, P_{jk}(\tau)p(\tau)e^{-s \tau}.
\label{eq:R_matrix_Laplace}
\end{equation}
In the previous equation, $R_{jk}(\tau)=P_{jk}(\tau)p(\tau)$ is the probability density that the system is reinitialized in the reset state $\ket{k}$ in the time interval $(\tau,\tau+\mbox{d}\tau)$ since the previous reset to the reset state $\ket{j}$. Note that $\widehat{R}_{jk}(s=0)$ is a Markov matrix since $\sum_{k}\widehat{R}_{jk}(s=0)=1$.
The resulting expression for $\widehat{\rho}(s)$ is (see Supplemental Material \cite{SM})  
\begin{equation}
\widehat{\rho}(s) = \widehat{\rho}_{01}(s) +\widehat{\rho}_{01}(s) \widehat{r}_{11}(s) +     \widehat{\rho}_{02}(s)\widehat{r}_{12}(s),
\label{eq:summation_density_resets_done}
\end{equation}
with 
\begin{equation}
\widehat{\rho}_{01}(s)=\int_{0}^{\infty} \mbox{d}\tau \, q(\tau) \rho_{\mathrm{free},1}(\tau) e^{-s\tau},
\label{eq:rho01_def}
\end{equation}
and analogously for $\widehat{\rho}_{02}(s)$ in terms of $\rho_{\mathrm{free},2}(t)$. In Eq.~\eqref{eq:rho01_def} we have introduced the \textit{survival probability} $q(\tau)=\int_{\tau}^{\infty}\mbox{d}\tau' \, p(\tau')$, which is the probability that no reset happens before time $\tau$. 

The case of Poissonian resetting with constant rate $\gamma$ corresponds to $p(\tau)=p_{\gamma}(\tau)=\gamma \, \mbox{exp}(-\gamma \tau)$ and $q(\tau)=q_{\gamma}(\tau)=\mbox{exp}(-\gamma \tau)$.  
The derivations in this manuscript apply, however, to arbitrary distributions $p(\tau)$.

The first term in Eq.~\eqref{eq:summation_density_resets_done}, involving $\widehat{\rho}_{01}(s)$, accounts for trajectories where no reset takes place and it is therefore given by the unitary time evolution from the initial state $\ket{\psi_0}=\ket{1}$ over the interval $[0,t]$. The coefficients $r_{11}(t) \geq 0$ and $r_{12}(t) \geq 0$, which are obtained from the inverse Laplace transform of $\widehat{r}_{11}(s)=\sum_{n=1}^{\infty}(\widehat{R}^n(s))_{11}$ and $\widehat{r}_{12}(s)=\sum_{n=1}^{\infty}(\widehat{R}^n(s))_{12}$ in Eq.~\eqref{eq:summation_density_resets_done}, can be interpreted as two \textit{effective time-dependent} rates of jumping into the reset states $\ket{1}$ and $\ket{2}$, respectively. Crucially, $r_{11}(t)$ and $r_{12}(t)$ are determined both by the resetting distribution $p(\tau)$ and by the Hamiltonian dynamics, which are coupled due to the conditional protocol.
The calculation of the NESS density matrix $\rho_{\mathrm{ness}}$ can then be directly performed from Eq.\eqref{eq:summation_density_resets_done} as $\rho_{\mathrm{ness}}=\lim_{t \to \infty} \rho(t)= \lim_{s \to 0} s \, \widehat{\rho}(s)$,
which gives:  
\begin{align}
\rho_{\mathrm{ness}} &= \frac{\widehat{R}_{21}(0)}{\widehat{R}_{21}(0)+\widehat{R}_{12}(0)}\frac{1}{\widehat{q}(0)}\int_{0}^{\infty}\mbox{d}\tau \, \rho_{\mathrm{free},1}(\tau) q(\tau) \nonumber \\
&+ \frac{\widehat{R}_{12}(0)}{\widehat{R}_{21}(0)+\widehat{R}_{12}(0)}\frac{1}{\widehat{q}(0)}\int_{0}^{\infty}\mbox{d}\tau \, \rho_{\mathrm{free},2}(\tau) q(\tau).    
\label{eq:NESS_reset_1}
\end{align}
This equation is the first main result of this work. It expresses $\rho_{\mathrm{ness}}$ as a statistical mixture of the unitary (reset-free) time evolution ensuing from the reset states $\ket{1}$ and $\ket{2}$. Fundamentally, both the weights of the terms in the sum couple the Hamiltonian dynamics with the reset via Eqs.~\eqref{eq:P_ij_Projector} and \eqref{eq:R_matrix_Laplace}. One can further check (see Supplemental Material \cite{SM}), as noted also in Ref.~\cite{Mukherjee2018} for resets to a single state, that $\rho_{\mathrm{ness}}$ has non-vanishing off-diagonal matrix elements in the basis of the eigenstates of the Hamiltonian $H$. This shows that $\rho_{\mathrm{ness}}$
is a genuinely non-equilibrium density matrix since it is different from the diagonal ensemble, which describes the relaxation at long times of closed quantum systems \cite{Mukherjee2018,ETH2016quantum}. 
\begin{figure*}[t]
    \centering
    \includegraphics[width=2\columnwidth]{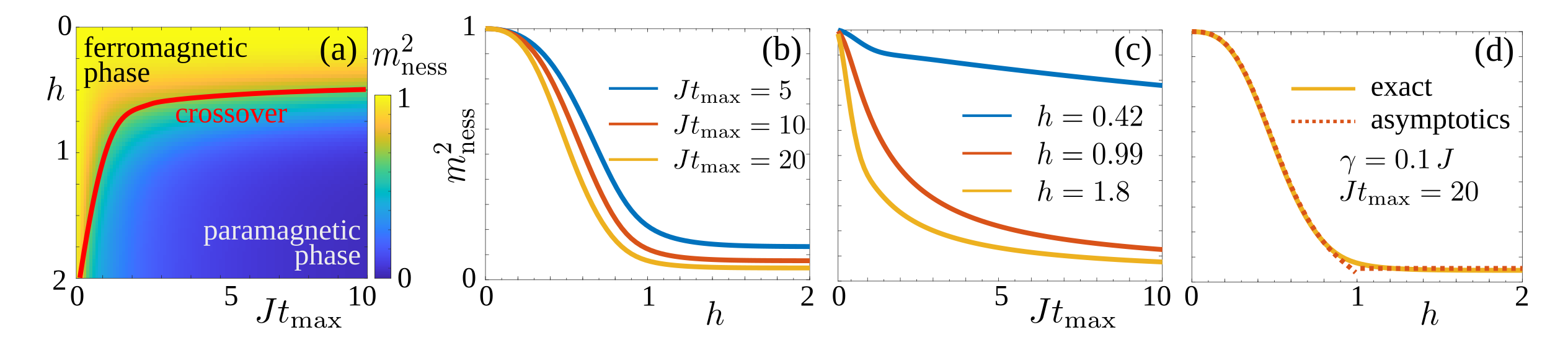} 
    \caption{\textbf{NESS phase diagram.} (a) Squared magnetization density, $m^2_{\mathrm{ness}}$, in the TFIC as a function of transverse field $h$ and (scaled) maximum reset time $J t_{\mathrm{max}}$. The red line is a guide for the eye showing the region where the crossover from the ferromagnetic to the paramagnetic stationary phase takes place. (b) Plot of $m^2_{\mathrm{ness}}$ as a function of $h$ for fixed values of $J t_{\mathrm{max}}=5,10,20$ (from top to bottom). As the maximum reset time increases the crossover becomes sharper and resembles more and more the equilibrium quantum phase transition behavior.  
    (c) Plot of $m^2_{\mathrm{ness}}$ as a function of $J t_{\mathrm{max}}$ for different fixed values of $h=0.42,0.99,1.8$ (from top to bottom). (d) Comparison between the result for $m^2_{\mathrm{ness}}$ as a function of $h$ from the exact evaluation of Eq.~\eqref{eq:squared_m_final} with $\Braket{\sigma^x(\tau)}$ obtained from the clustering property of the two-point function (yellow solid line), and the result obtained using the asymptotic expression of $\Braket{\sigma^x(\tau)}$ valid for large $\tau$ (red dashed line). In the Figure $\gamma=0.1 J$.}
    \label{fig:2}
\end{figure*}

A complementary way to the renewal equation approach, is to construct an evolution equation for $\rho(t)$, which for classical systems has been done in Ref.~\cite{eule2016non}. This perspective is useful as it sheds light on the effective open system dynamics which emerges upon superimposing resetting to the unitary time evolution of the system. For our conditional reset protocol we find (see Supplemental Material \cite{SM}) from Eq.~\eqref{eq:summation_density_resets_done} the \textit{generalized Lindblad equation}
\begin{eqnarray}
\frac{\mbox{d}\rho(t)}{\mbox{d}t} &=& \mathcal{L}[\rho(t)]+r_{11}(t)\ket{1}\bra{1}+r_{12}(t)\ket{2}\bra{2} \nonumber \\
&&-\int_{0}^{t} \mbox{d}t' r(t-t') e^{(t-t')\mathcal{L}} \rho(t')
\label{eq:generalized_master_equation}, 
\end{eqnarray}
with the time-dependent rates $r_{1k}(t)$, with $k=1,2$, written in terms of the density matrix (see Supplemental Material \cite{SM}) as
\begin{equation}
r_{1k}(t) = \int_{0}^{t}\mbox{d}t' r(t-t')\mbox{Tr}\left[\mathcal{P}_k e^{(t-t')\mathcal{L}}\rho(t')\right].
\label{eq:non_Markov_rates}
\end{equation}
In Eq.~\eqref{eq:generalized_master_equation}, $\mathcal{L}[\rho]=-i[H,\rho]$ is the generator of the unitary time evolution and $r(t)$ satisfies $r_{11}(t)+r_{12}(t)=\int_{0}^{t}\mbox{d}t' r(t')$, which ensures a trace-preserving dynamics. 

Equations \eqref{eq:generalized_master_equation} and \eqref{eq:non_Markov_rates} are our second main result. They show that stochastic resetting is generating an engineered dissipative process which leads to an emergent open dynamics for the system without an actual environment being present.
The dynamics in Eqs.~\eqref{eq:generalized_master_equation} and \eqref{eq:non_Markov_rates} is in general non-Markovian because of the presence of the integral over the complete time history of the process. This is a
consequence, as in the classical case of Ref.~\cite{eule2016non}, of the fact
that for non-Poissonian distributions $p(\tau)$ one has to keep
track of the time elapsed since the last reset. Only in the
Poissonian case, $p_{\gamma}(\tau)$, resetting occurs at constant rate $\gamma$,
independently of the time elapsed since the last reset. In
this case one verifies that $r(t)=\gamma \delta(t)$ in Eq.~\eqref{eq:generalized_master_equation}, which gives back a Markovian-Lindblad dynamics \cite{SM,Hartmann2006,Linden2010,Armin2020,Rose2018spectral,riera2020measurement}.

\emph{The quantum Ising chain in a transverse field} We particularize in this Section our theory to the TFIC introduced above [see Eq.~\eqref{eq:TFIC_Hamiltonian}]. Periodic boundary conditions will be henceforth assumed. 
The quantum critical point at $h=h_c=1$ separates \cite{lieb1961two,pikin1966thermodynamics,pfeuty1970one,kogut1979}, at zero temperature and in the thermodynamic limit $N \to \infty$, the paramagnetic ($h>1$) from the ferromagnetic ($h<1$) phase. In the latter, because of the spontaneous breaking of the $\mathbb{Z}_2$ symmetry, the ground-state expectation value of the magnetization density $m$ becomes non-zero $\tensor*[_{\pm}]{\Braket{\mbox{GS}(h)|m|\mbox{GS}(h)}}{_{\pm}}=\pm (1-h^2)^{1/8}$, 
where $\ket{\mbox{GS}(h)}_{\pm}$ are the two degenerate ground states for $h<1$.

Because of the $\mathbb{Z}_2$ symmetry, the matrices $P_{jk}(\tau)$ and $\widehat{R}_{jk}(s)$ in Eqs.~\eqref{eq:P_ij_Projector} and \eqref{eq:R_matrix_Laplace} are symmetric. This, in turn, implies that the magnetization density in the NESS in Eq.~\eqref{eq:NESS_reset_1} is identically zero: $\mbox{Tr} (m \, \rho_{\mathrm{ness}})= 0$ (and the same for all the odd operators under the $\mathbb{Z}_2$ symmetry; see Supplemental Material \cite{SM}). The natural order parameter to distinguish between the ordered and the disordered phase is then the squared magnetization density $m^2_{\mathrm{ness}}=\mbox{Tr} (m^2 \rho_{\mathrm{ness}})$. For the latter one can derive, in the thermodynamic limit $N \to \infty$, the expression (see Supplemental Material \cite{SM})
\begin{equation}
m^2_{\mathrm{ness}}=
 \frac{1}{\widehat{q}(0)}\int_{0}^{\infty}\mbox{d}\tau \, q(\tau) \Braket{\sigma^x(\tau)}^2,
\label{eq:squared_m_final} 
\end{equation}
which directly connects the NESS expectation value of $m^2_{\mathrm{ness}}$ in the presence of reset with the magnetization density quench dynamics $\Braket{m(\tau)}=\Braket{\sigma^x(\tau)}$ (with the last equality coming from the choice of periodic boundary conditions). Equation \eqref{eq:squared_m_final} can be evaluated exactly in the thermodynamic limit upon extracting $\Braket{\sigma^x(\tau)}$ from the cluster decomposition principle of the two-point function $\Braket{\sigma^x_n \sigma^x_{n+l}}(\tau)$ at large separation $l$ (see Supplemental Material \cite{SM}), as outlined first in \cite{McCoy1,McCoy2,McCoy3,McCoy4} for the equilibrium correlation function and in \cite{calabrese2011quantum,calabrese2012quantum1,calabrese2012quantum2} for the non-equilibrium quench dynamics. In the latter case, it is known \cite{calabrese2011quantum,calabrese2012quantum1,calabrese2012quantum2,refereeeIsing2021} that $\Braket{\sigma^x(\tau)} \propto \mbox{exp}(-t \Lambda(h,h_0))$ relaxes exponentially to zero at long times with a characteristic time scale $\Lambda(h,h_0)^{-1}>0$ set by the quench (see Supplemental Material \cite{SM}).   

In Fig.~\ref{fig:2} we evaluate Eq.~\eqref{eq:squared_m_final} for the waiting time distribution $p(\tau)=p_{\gamma,t_{\mathrm{max}}}(\tau)= \gamma e^{-\gamma \tau}/(1-e^{-\gamma t_{\mathrm{max}}})\Theta(t_{\mathrm{max}}-\tau)$, which represents a ``chopped Poissonian distribution'' with rate $\gamma$ and maximum reset time $t_{\mathrm{max}}$ [$\Theta(\tau)$ denotes the Heaviside step function]. The NESS reflects the equilibrium quantum phase transition through a sharp crossover due to the competition between the maximum reset time $t_{\mathrm{max}}$ and the order parameter decay time $\Lambda(h,h_0)^{-1}$. Upon increasing $J t_{\mathrm{max}}$ with fixed $\gamma/J$ (or, vice-versa, decreasing $\gamma/J$ with $J t_{\mathrm{max}}$ fixed) the crossover becomes very sharp, as shown in Fig.~\ref{fig:2}(b) for $\gamma=0.1 J$ and $J t_{\mathrm{max}}=20$. We emphasize that $m^2_{\mathrm{ness}}$ is always different from zero, even for large values of $h$ deep in the paramagnetic phase, for any finite value of $J t_{\mathrm{max}}$ or $\gamma/J$. This means that $\rho_{\mathrm{ness}}$, thanks to the resetting protocol, retains part of the long-range order stored in the reset states $\ket{1}$ and $\ket{2}$. For the reset-free quench time evolution, indeed, the diagonal ensemble 
predicts a vanishing expectation value of $m^2$ and therefore a complete loss of the order of $\ket{1}$ and $\ket{2}$. As $Jt_{\mathrm{max}}\to 0$ one encounters the quantum Zeno effect \cite{zeno1977one,zeno1977two} since the initial state does not evolve in time due to very frequent resets and therefore the squared magnetization density remains close to $1$.
In the limit of large $J t_{\mathrm{max}}$ (small $\gamma/J$), when the crossover gets sharp and it closely resembles the equilibrium quantum phase transition, a large time, on average, elapses between consecutive resets. As a consequence, one can evaluate Eq.~\eqref{eq:squared_m_final} using the asymptotics of $\Braket{\sigma^x(\tau)}\propto \mbox{exp}(-t \Lambda(h,h_0))$ for long times $\tau \to \infty$, which has been analytically derived in Refs.~\cite{calabrese2011quantum,calabrese2012quantum1,calabrese2012quantum2,SM}. The result is shown in Fig.~\ref{fig:2}(d) (see Supplemental Material \cite{SM}). We can see that the asymptotics of the order parameter is in excellent agreement with the exact numerical evaluation of Eq.~\eqref{eq:squared_m_final}. Only for values of $h \gtrsim 0.9$ discrepancies are visible. This is caused by the fact that when $h$ becomes large the order parameter $\Braket{\sigma^x(\tau)}$ quickly relaxes to zero as a function of $\tau$ and the integral in Eq.~\eqref{eq:squared_m_final} is therefore no longer dominated by the behavior of $\Braket{\sigma^x(\tau)}$ at large times.  

\emph{Outlook.} We have presented a general protocol for constructing the NESS of a unitary many-body dynamics interspersed by random resets. Using the TFIC as exemplary case, we have shown that this NESS is intimately connected to the dynamics following a quantum quench, and that it embodies the ground-state quantum phase transition of the TFIC through a sharp crossover. NESS with sharp crossovers can here be exploited to design collectively enhanced sensing protocols where a small change of the parameter (external perturbation) driving the quantum phase transition translates into an easy-to-detect macroscopic response, as, e.g., discussed in Refs. \cite{raghunandan2018,Thzdetector}. A possible advantage of our protocol is, however, that the establishment of the NESS relies on stochastic resetting, which diminishes the impact of uncontrolled dissipative effects such as heating.

\emph{Acknowledgements.} We  acknowledge support  from the  “Wissenschaftler R\"uckkehrprogramm GSO/CZS” of the Carl-Zeiss-Stiftung and the German  Scholars Organization e.V., from the European Union's Horizon 2020 research and innovation program under Grant Aagreement No.~800942 (ErBeStA), as well as from the Baden-W\"urttemberg Stiftung through Project No.~BWST\_ISF2019-23.
\bibliography{biblioPRB}

\end{document}



\title{ Supplemental Material \\ {Designing} non-equilibrium states of quantum matter through stochastic resetting \\  }

\author{Gabriele Perfetto}
\affiliation{Institut f\"ur Theoretische Physik, Eberhard Karls Universit\"at T\"ubingen, Auf der
Morgenstelle 14, 72076 T\"ubingen, Germany.}
\author{Federico Carollo}
\affiliation{Institut f\"ur Theoretische Physik, Eberhard Karls Universit\"at T\"ubingen, Auf der
Morgenstelle 14, 72076 T\"ubingen, Germany.}
\author{Matteo Magoni}
\affiliation{Institut f\"ur Theoretische Physik, Eberhard Karls Universit\"at T\"ubingen, Auf der
Morgenstelle 14, 72076 T\"ubingen, Germany.}
\author{Igor Lesanovsky}
\affiliation{Institut f\"ur Theoretische Physik, Eberhard Karls Universit\"at T\"ubingen, Auf der
Morgenstelle 14, 72076 T\"ubingen, Germany.}
\affiliation{School of Physics and Astronomy and Centre for the Mathematics and Theoretical
Physics of Quantum Non-Equilibrium Systems, The University of Nottingham, Nottingham,
NG7 2RD, United Kingdom.}
\maketitle

\setcounter{equation}{0}
\setcounter{figure}{0}
\setcounter{table}{0}
\setcounter{page}{1}
\makeatletter
\renewcommand{\theequation}{S\arabic{equation}}
\renewcommand{\thefigure}{S\arabic{figure}}
\renewcommand{\bibnumfmt}[1]{[S#1]}
\renewcommand{\citenumfont}[1]{S#1}

{Here} we derive in detail the results presented in the main text. The Supplemental Material is organized as follows:  
in Section \ref{sec:ness} we show how the conditional resetting dynamics can be formulated in full generality using the formalism of semi-Markov processes \cite{janssen2006applied}; in Section \ref{sec:TFIC} we particularize the analysis to the quantum Ising chain in a transverse field (TFIC), for which we show how to compute the phase diagram of the magnetization density squared $m^2_{\mathrm{ness}}$ in the non-equilibrium steady state (NESS).

\section{Conditional reset in isolated many-body quantum systems}
\label{sec:ness}
In this Section we provide the general derivation of Eqs.~(3)-(8) of the main text. In particular, in Subsec.~\ref{sec:ness_renewal} we show how the NESS density matrix $\rho_{\mathrm{ness}}$ in Eq.~(6) of the main text can be analytically constructed using the renewal equation approach. In Subsec.~\ref{sec:ness_master_eq} we follow a complementary approach based on the corresponding evolution equation for the density matrix $\rho(t)$ as a function of time in the presence of reset. In particular, we obtain the generalized non-Markovian Lindblad equation in Eq.~(7) of the main text. 

\subsection{Renewal equation approach and construction of the NESS}
\label{sec:ness_renewal}
We consider a many-body quantum system composed by $N$ particles/spins, which undergoes unitary time evolution according to a certain Hamiltonian $H$.
The Hamiltonian in this Section will be left general. In Sec.~\ref{sec:TFIC} we will consider the specific case of the TFIC Hamiltonian. The unitary time evolution is interrupted by measurements of some observables. The time between consecutive measurements is distributed according to the waiting time probability density $p(\tau)$, i.e., after a measurement the next measurement occurs in the time interval $(\tau, \tau +\mbox{d} \tau)$ with probability $p(\tau)\mbox{d}\tau$. The waiting time probability density is normalized to one  
\begin{equation}
\int_{0}^{\infty} p(\tau) \mbox{d}\tau = 1.
\label{eq:normalization_p}
\end{equation}
The following derivation applies for arbitrary $p(\tau)$. To account, however, for the finite coherence time attained in cold-atom systems, we will be particularly interested in the ``chopped exponential distribution'' $p_{\gamma,t_{\mathrm{max}}}(\tau)$ with rate $\gamma$ and maximum reset time $t_{\mathrm{max}}$, which we define as
\begin{equation}
p(\tau) = p_{\gamma,t_{\mathrm{max}}}(\tau) = \frac{\gamma}{1-\mbox{exp}(-\gamma t_{\mathrm{max}})} e^{-\gamma \tau} \Theta(t_{\mathrm{max}}-\tau), \, \, \, \mbox{with} \quad 0<\tau<\infty,
\label{eq:chopped_exp}
\end{equation}
with $\Theta$ denoting the Heaviside step function.
It is useful to consider the two limiting cases of Eq.~\eqref{eq:chopped_exp} for $t_{\mathrm{max}} \to \infty$ (with fixed $\gamma$) and for $\gamma \to 0$ (for fixed $t_{\mathrm{max}}$). In the former case we have that $p_{\gamma,t_{\mathrm{max}}}(\tau)$ reduces to the Poisson (exponential) distribution: $p_{\gamma}(\tau)$
\begin{equation}
\lim_{t_{\mathrm{max}}\to \infty}p_{\gamma,t_{\mathrm{max}}}(\tau)=p_{\gamma}(\tau)=\gamma \, e^{-\gamma \tau}.    
\label{eq:exponential_distribution}
\end{equation}
In the limit $\gamma \rightarrow 0$, on the other hand, $p_{\gamma,t_{\mathrm{max}}}(\tau)$ reduces to a uniform distribution $p_{t_{\mathrm{max}}}(\tau)$ in the interval $(0,t_{\mathrm{max}})$:
\be
\lim_{\gamma \rightarrow 0} p_{\gamma,t_\mathrm{max}}(\tau)=    p_{t_\mathrm{max}}(\tau) = \frac{1}{t_{\mathrm{max}}} \, \Theta(t_{\mathrm{max}}-\tau),  \, \, \, \mbox{with} \quad 0<\tau<\infty.
\label{eq:uniform_waiting_time}
\ee
We define for future convenience the survival probability $q(\tau)$
\begin{equation}
q(\tau) = \int_{\tau}^{\infty} \mbox{d}\tau' p(\tau'), \label{eq:survival_probability}   
\end{equation}
which gives the probability that no measurement has been performed before time $\tau$. In the case of $p_{\gamma,t_{\mathrm{max}}}$ one has
\begin{equation}
q_{\gamma,t_{\mathrm{max}}}(\tau) = \int_{\tau}^{\infty} \mbox{d}\tau' p_{\gamma,t_{\mathrm{max}}}(\tau') = \left(\frac{e^{-\gamma\tau}-e^{-\gamma t_{\mathrm{max}}}}{1-e^{-\gamma t_{\mathrm{max}}}}\right)\Theta(t_{\mathrm{max}}-\tau), \, \, \, \mbox{with} \quad 0<\tau<\infty.
\label{eq:survival_chop_exp}
\end{equation}
In the limit $t_{\mathrm{max}} \to \infty$ from Eq.~\eqref{eq:exponential_distribution} one has
\begin{equation}
\lim_{t_{\mathrm{max}} \to \infty} q_{\gamma,t_{\mathrm{max}}}(\tau) = q_{\gamma}(\tau) = e^{-\gamma \tau},
\label{eq:exponential_survival}
\end{equation}
while in the limit $\gamma \rightarrow 0$ Eq.~\eqref{eq:uniform_waiting_time} gives
\be
\lim_{\gamma \rightarrow 0} q_{\gamma,t_\mathrm{max}}(\tau)=    q_{t_\mathrm{max}}(\tau) = \frac{t_{\mathrm{max}}-\tau}{t_{\mathrm{max}}}\Theta(t_{\mathrm{max}}-\tau) , \, \, \, \mbox{with} \quad 0<\tau<\infty.
\label{eq:survival_uniform}
\ee

When a measurement is performed, the many-body wavefunction of the system is \textit{reset} to a state chosen within a set of ``reset states''. We consider in this Section, for concreteness, the case of a set of reset states composed by two elements. As we will comment later on, the generalization to an arbitrary number of reset states is straightforward and it does not bear additional conceptual difficulties. We consider the two following reset states
\begin{subequations}
\begin{align}
\ket{1} &\equiv \ket{\uparrow, \uparrow \dots \uparrow}, \\ 
\ket{2} &\equiv  \ket{\downarrow, \downarrow \dots \downarrow}.
\end{align}
\label{eq:reset_states}%
\end{subequations}
We denote with $\ket{\uparrow} = \frac{1}{\sqrt{2}}(1,1)$ and $\ket{\downarrow}=\frac{1}{\sqrt{2}}(1,-1)$ the eigenstates of the Pauli matrix $\sigma^x$. Both the states in Eq.~\eqref{eq:reset_states} present ferromagnetic long-range order and, indeed, they are the two degenerate ground states of the TFIC Hamiltonian in Eq.~\eqref{eq:TFIC_Hamiltonian} for $h=0$, as we shall see in Sec.~\ref{sec:TFIC}. We then define the sets 
\begin{subequations}
\begin{align}
C^0 &= \left\{\ket{C_{\mu}^0} : \frac{1}{N}\sum_{n=1}^N \sigma^x_n \ket{C_{\mu}^0} = 0, m_{\mu}^0 =0, \quad \mu=1,2 \dots \, \, \frac{N!}{((N/2)!)^2} \right\}, 
\\ 
C^1 &= \left\{\ket{C_{\mu}^1} : \frac{1}{N}\sum_{n=1}^N \sigma^x_n \ket{C_{\mu}^1} = m_{\mu}^1  \ket{C^1_{\mu}}, m_{\mu}^1 >0, \quad \mu=1,2 \dots \, \, 2^{N-1}-\frac{N!}{2 ((N/2)!)^2} \right\}, 
\\
C^2 &= \left\{\ket{C_{\mu}^2} : \frac{1}{N}\sum_{n=1}^N \sigma^x_n \ket{C_{\mu}^2} = m_{\mu}^2  \ket{C_{\mu}^2}, m_{\mu}^2 <0, \quad \quad \mu=1,2 \dots \, \, 2^{N-1}-\frac{N!}{2 ((N/2)!)^2} \right\},  
\end{align}
\label{eq:reset_sets}%
\end{subequations}
which are nothing but the sets of eigenstates of the magnetization operator $\sum_{n=1}^N \sigma^x_n/N$ with zero, positive and negative eigenvalues, respectively. Note that for an odd number $N$ of spins $C^0$ is just the empty set. The sets $C^0$, $C^1$ and $C^2$ define a partition of the whole Hilbert space $\mathcal{H}$ with $C^0 \cup C^1 \cup C^2$ defining a complete basis of $\mathcal{H}$ and $C^0 \cap C^1 = C^1 \cap C^2= C^0 \cap C^2= \emptyset$. In addition, the orthonormality condition
\be
\Braket{C_{\mu}^j|C_{\mu'}^k}=\delta_{j,k}\delta_{\mu,\mu'},
\label{eq:orthonormality_reset_sets_elements}
\ee
is satisfied by the elements of $C^0$, $C^1$ and $C^2$. Clearly $\ket{1} \in C^1$, while $\ket{2} \in C^2$.
When a reset event happens, the system is re-initialized in one of the two states in Eq.~\eqref{eq:reset_states} according to the probability
\begin{equation}
P_{jk}(\tau) = \sum_{\mu} |\Braket{C_{\mu}^k|\mbox{exp}(-i H \tau) |j}|^2 + \frac{1}{2}\sum_{\mu} |\Braket{C_{\mu}^0|\mbox{exp}(-i H \tau) |j}|^2  = \Braket{\psi_j(\tau)| \mathcal{P}_k | \psi_j(\tau)},
\label{eq:probability_two_resets}
\end{equation}
with $j,k=1,2$ and the summation in $\mu$ running over the elements of the set corresponding to the index $k$ or $0$ according to Eq.~\eqref{eq:reset_sets}. In Eq.~\eqref{eq:probability_two_resets} we have defined the operator $\mathcal{P}_k$ corresponding to the set $C^k$ as
\begin{equation}
\mathcal{P}_k =  \sum_{\mu} \ket{C^k_{\mu}} \bra{C^k_{\mu}} +\frac{1}{2} \sum_{\mu} \ket{C^0_{\mu}} \bra{C^0_{\mu}} 
\label{eq:projector},
\end{equation}
and the unitary evolution from the state $\ket{j}$
\begin{equation}
\ket{\psi_j(\tau)} = \mbox{exp}(-i H \tau) \ket{j},  \, \, \, \mbox{with} \, \, j=1,2.
\label{eq:unitary_evolution_states}
\end{equation}
Note that 
\begin{equation}
\sum_{k=1}^2 P_{jk}(\tau) = \Braket{\psi_j(\tau)| (\mathcal{P}_1 + \mathcal{P}_2)  | \psi_j(\tau)} = \Braket{\psi_j(\tau)| \psi_j(\tau)} =1, 
\label{eq:normalization_Pij}
\end{equation}
since $C^0 \cup C^1 \cup C^2$ is a complete basis of the Hilbert space $\mathcal{H}$, as already commented after Eq.~\eqref{eq:reset_sets}. On the basis of the wavefunction interpretation of quantum mechanics \cite{sakurai2006advanced}, we, therefore, see that $P_{jk}(\tau)$ in Eq.~\eqref{eq:probability_two_resets} represents the probability of measuring a non-negative ($k=1$) or non-positive ($k=2$) value of the magnetization after a time $\tau$ since the previous reset towards the reset state $j$. The factor $1/2$ in front of the second summation of Eqs.~\eqref{eq:probability_two_resets} and \eqref{eq:projector} is chosen such that if a zero magnetization is measured, the many-body state is reset with probability $1/2$ to the state $\ket{1}$ and with probability $1/2$ to the state $\ket{2}$. We also remark that the operator $\mathcal{P}_k$ in Eq.~\eqref{eq:projector} is not a projector. This, however, is not necessary for the following derivation to hold. Indeed, only the normalization condition for $P_{jk}(\tau)$ in Eq.~\eqref{eq:normalization_Pij} is necessary for the semi-Markov construction to be well-defined.
From Eq.~\eqref{eq:probability_two_resets}, as a matter of fact, one realizes that the outcome of a reset transition depends on the time $\tau$ elapsed since the last reset and on the outcome $j$ of the previous reset, but, fundamentally, not on the previous history of the dynamics. This kind of dynamics precisely defines a \textit{semi-Markov process} (see, e.g., Ref.~\onlinecite{janssen2006applied}). The semi-Markov structure induced by Eqs.~\eqref{eq:probability_two_resets}-\eqref{eq:normalization_Pij} is the essential ingredient allowing for an exact, analytical, construction of the NESS, as we now show. 
We combine the waiting time probability density $p(\tau)$ and the probability $P_{jk}(\tau)$ in Eq.~\eqref{eq:probability_two_resets} into the following probability
\begin{equation}
R_{jk}(\tau)= P_{jk}(\tau) p(\tau), \label{eq:h_probability}
\end{equation}
which satisfies, for any value of $j$, the following normalization condition from Eqs.~\eqref{eq:normalization_p} and \eqref{eq:normalization_Pij}
\begin{equation}
\sum_{k=1}^{2} \int_0^\infty \mbox{d}\tau R_{jk}(\tau) =1.
\label{eq:normalization_h}
\end{equation}
Without loss of generality, we take the initial state of the system at time $t=0$ to be the reset state $\ket{1}$ in Eq.~\eqref{eq:reset_states}. 
\begin{equation}
\ket{\psi(0)} = \ket{1}, \,\,\, \rho(0)=\ket{\psi(0)}\bra{\psi(0)}=\ket{1}\bra{1}. \label{eq:initial_state}    
\end{equation}
We emphasize, however, that the specific choice of the initial state $\ket{\psi_0}$ influences only the finite time, transient, dynamics.
It is, indeed, straightforward to generalize the present derivation to an arbitrary initial state $\ket{\psi(0)}$ and, thereby, to show that the steady state density matrix $\rho_{\mathrm{ness}}$, in Eq.(6) of the main text, does not depend on the choice of $\ket{\psi(0)}$. 

We have now defined all the quantities necessary for the study of the state-dependent reset dynamics. One is now interested in deriving an equation for the density matrix $\rho(t)$ under such kind of dynamics.
To do this we consider the observation time $t$ to be fixed and we write the density matrix $\rho_n(t)$ at time $t$ assuming that exactly $n\geq 0$ resets happened in the interval $(0,t)$.
We write the first terms to provide intuition on the emerging general structure:
\begin{subequations}
\begin{align}
&\rho_0(t) = q(t) \rho_{\mathrm{free},1}(t), \, \, \, n=0,  \\
&\rho_1(t) = \int_{0}^{t} \mbox{d}\tau_1 \, \left[ R_{11}(\tau_1)q(t-\tau_1)\rho_{\mathrm{free},1}(t-\tau_1)+R_{12}(\tau_1)q(t-\tau_1)\rho_{\mathrm{free},2}(t-\tau_1)\right], \, \, \, n=1 \\
&\rho_2(t) = \int_{0}^{t} \mbox{d}\tau_1 \int_{0}^{t-\tau_1} \mbox{d}\tau_2 \, \Big[ q(t-\tau_1-\tau_2) \rho_{\mathrm{free},1}(t-\tau_1-\tau_2)(R_{12}(\tau_1)R_{21}(\tau_2)+R_{11}(\tau_1)R_{11}(\tau_2))+  \\ 
&\qquad \qquad \qquad \qquad \quad  \quad  \quad +q(t-\tau_1-\tau_2) \rho_{\mathrm{free},2}(t-\tau_1-\tau_2)(R_{11}(\tau_1)R_{12}(\tau_2)+R_{12}(\tau_1)R_{22}(\tau_2))\Big], \, \, \, n=2 \dots  \\
&\rho_n(t) = \int_{0}^{t} \mbox{d}\tau_1 \int_{0}^{t-\tau_1} \mbox{d}\tau_2 \dots \int_{0}^{t-\tau_1-\tau_2 \dots -\tau_{n-1}} \mbox{d}\tau_{n} \, \Big[ \nonumber \\
& \, \, \, \qquad \qquad \qquad \qquad q(t-\tau_1-\tau_2 \dots -\tau_n) \rho_{\mathrm{free},1}(t-\tau_1-\tau_2 \dots -\tau_n) \Big(\sum_{i_1,i_2,\dots i_{n-1}}R_{1i_1}(\tau_1)R_{i1,i2}(\tau_2)\dots R_{i_{n-1}1}(\tau_n)\Big)+  \\ 
& \, \, \, \qquad \qquad \qquad \qquad q(t-\tau_1-\tau_2 \dots -\tau_n) \rho_{\mathrm{free},2}(t-\tau_1-\tau_2 \dots -\tau_n) \Big(\sum_{i_1,i_2,\dots i_{n-1}}R_{1i_1}(\tau_1)R_{i1,i2}(\tau_2)\dots R_{i_{n-1}2}(\tau_n)\Big)\Big],
\end{align}
\label{eq:reset_expansion}%
\end{subequations}
with 
\begin{equation}
\rho(t) = \sum_{n=0}^{\infty} \rho_n(t), 
\label{eq:full_density_matrix_reset}
\end{equation}
and the definition from Eq.~\eqref{eq:unitary_evolution_states}
\begin{equation}
\rho_{\mathrm{free},j}(t)=\ket{\psi_j(t)}\bra{\psi_j(t)}=\mbox{exp}(-iH t) \ket{j}\bra{j} \mbox{exp}(iHt),  \, \, \, \mbox{with} \, \, j=1,2, \label{eq:unitary_evolution_density_matrix}    
\end{equation}
the reset-free, unitary, evolution starting from the state $\ket{j}$. 
The physical interpretation of Eq.~\eqref{eq:reset_expansion} is clear. The first equation for $\rho_0(t)$ simply accounts for the case of no reset up to time $t$ and therefore it is analogous to the reset-free evolution weighted by the survival probability $q(t)$. The second equation for $\rho_1(t)$ accounts for the case of only one reset happening at time $\tau_1 \in (0,t)$ with probability $p(\tau_1)$. For the remaining time $t-\tau_1$ the system evolves without additional resetting and therefore the weight $q(t-\tau_1)$ is included.
The convolution structure in Eq.~\eqref{eq:reset_expansion} suggests using the Laplace transform, which is defined for an arbitrary function $f(t)$ of time $t$ as: 
\begin{equation}
\widehat{f}(s) = \int_{0}^{\infty} \mbox{d}t \,  f(t) e^{-st}. 
\label{eq:Laplace_definition}    
\end{equation}
Taking the Laplace transform of $\rho_n(t)$ in Eq.~\eqref{eq:reset_expansion} one obtains 
\begin{equation}
\widehat{\rho}_n(s) = \widehat{\rho}_{01}(s) \Big(\sum_{i_1,i_2,\dots i_{n-1}}\widehat{R}_{1i_1}(s)\widehat{R}_{i1,i2}(s)\dots \widehat{R}_{i_{n-1}1}(s)\Big)+\widehat{\rho}_{02}(s) \Big(\sum_{i_1,i_2,\dots i_{n-1}}\widehat{R}_{1i_1}(s)\widehat{R}_{i1,i2}(s)\dots \widehat{R}_{i_{n-1}2}(s)\Big),
\label{eq:intermediate_step_summation}
\end{equation}
with the definition (cf., Eq.~\eqref{eq:unitary_evolution_density_matrix})
\begin{equation}
\widehat{\rho}_{01}(s) = \widehat{\rho}_0(s)=\int_{0}^{\infty} \mbox{d}t \, q(t) \rho_{\mathrm{free},1}(t) e^{-st}, \quad \mbox{and} \quad \widehat{\rho}_{02}(s) = \int_{0}^{\infty} \mbox{d}t \, q(t) \rho_{\mathrm{free},2}(t) e^{-st},   
\label{eq:rho_up_down_def}
\end{equation}
and the matrix $\widehat{R}_{jk}(s)$ the Laplace transform of the matrix $R_{jk}(\tau)$ in Eq.~\eqref{eq:h_probability} (with the same notation as in Eq.(3) of the main text) 
\begin{equation}
\widehat{R}(s) =\begin{pmatrix}
\widehat{R}_{11}(s) & \widehat{R}_{12}(s) \\ 
\widehat{R}_{21}(s) & \widehat{R}_{22}(s) 
\end{pmatrix}, \quad \widehat{R}_{jk}(s) = \int_{0}^{\infty} \mbox{d}\tau P_{jk}(\tau) p(\tau) e^{-s\tau}, \, \, \, \mbox{with} \,\,\, j,k =1,2.
\label{eq:R_matrix}
\end{equation}
The matrix $\widehat{R}_{jk}(s) \geq 0$ is non-negative. From Eq.~\eqref{eq:intermediate_step_summation} in Laplace space one then has 
\begin{equation}
\widehat{\rho}(s) = \widehat{\rho}_{01}(s) +\widehat{\rho}_{01}(s) \sum_{n=1}^{\infty} \left(\widehat{R}^n(s)\right)_{11} +     \widehat{\rho}_{02}(s) \sum_{n=1}^{\infty} \left(\widehat{R}^n(s)\right)_{12},
\label{eq:summation_density_resets}
\end{equation}
which comes from recognizing the matrix product structure in Eq.~\eqref{eq:reset_expansion}. 
In the Laplace space the matrix $\widehat{R}(s)$ satisfies the following relations for every row $j$
\begin{align}
\sum_{k} \widehat{R}_{jk}(s) <1 \, \, \, \mbox{for} \, \, s>0, \qquad 
\sum_{k} \widehat{R}_{jk}(s) >1 \, \, \, \mbox{for} \, \, s<0, \qquad  
\sum_{k} \widehat{R}_{jk}(s) =1 \, \, \, \mbox{for} \, \, s=0.
\label{eq:matrix_R_norm}
\end{align}
From the last equality one recognizes that for $s=0$ the matrix $\widehat{R}(s)$ is a stochastic-Markovian matrix  \cite{janssen2006applied}. The geometric sum in Eq.~\eqref{eq:summation_density_resets} is convergent provided the spectral radius, i.e., the absolute value of the largest eigenvalue of $\widehat{R}(s)$, is smaller than $1$. From Eq.~\eqref{eq:matrix_R_norm}\cite{berman1994nonnegative,yang2011simple}, we see that this is true for $s>0$: 
\begin{equation}
\sum_{n=1}^{\infty} \widehat{R}^n(s) = (\mathbb{I}-\widehat{R}(s))^{-1} -\mathbb{I}, \quad \mbox{for} \, \, s>0.
\label{eq:sum_geometric_matrix}
\end{equation}
We can explicitly invert the matrix $\widehat{R}(s)$ since it is of dimension $2 \times 2$
\begin{subequations}
\begin{align}
\widehat{r}_{11}(s) \equiv ((\mathbb{I}-\widehat{R}(s))^{-1} -\mathbb{I})_{11}= \frac{\widehat{R}_{11}(s)(1-\widehat{R}_{22}(s))+\widehat{R}_{21}(s)\widehat{R}_{12}(s)}{(1-\widehat{R}_{11}(s))(1-\widehat{R}_{22}(s))-\widehat{R}_{12}(s)\widehat{R}_{21}(s)}, \\
\widehat{r}_{12}(s) \equiv ((\mathbb{I}-\widehat{R}(s))^{-1} -\mathbb{I})_{12}= \frac{\widehat{R}_{12}(s)}{(1-\widehat{R}_{11}(s))(1-\widehat{R}_{22}(s))-\widehat{R}_{12}(s)\widehat{R}_{21}(s)}, \\
\widehat{r}_{21}(s) \equiv ((\mathbb{I}-\widehat{R}(s))^{-1} -\mathbb{I})_{21}= \frac{\widehat{R}_{21}(s)}{(1-\widehat{R}_{11}(s))(1-\widehat{R}_{22}(s))-\widehat{R}_{12}(s)\widehat{R}_{21}(s)}, \\
\widehat{r}_{22}(s) \equiv ((\mathbb{I}-\widehat{R}(s))^{-1} -\mathbb{I})_{22}= \frac{\widehat{R}_{22}(s)(1-\widehat{R}_{11}(s))+\widehat{R}_{21}(s)\widehat{R}_{12}(s)}{(1-\widehat{R}_{11}(s))(1-\widehat{R}_{22}(s))-\widehat{R}_{12}(s)\widehat{R}_{21}(s)}.
\end{align}
\label{eq:rates}%
\end{subequations}
The inverse Laplace transform  $r_{11}(t),r_{12}(t),r_{21}(t)$ and $r_{22}(t)$ of the previous equations can be performed in general numerically. The expression for $\widehat{r}_{11}(s)$ and $\widehat{r}_{12}(s)$ can then be inserted into Eq.~\eqref{eq:summation_density_resets} attaining
\begin{equation}
\widehat{\rho}(s) = \widehat{\rho}_{01}(s) +\widehat{\rho}_{01}(s) \widehat{r}_{11}(s) +     \widehat{\rho}_{02}(s)\widehat{r}_{12}(s),
\label{eq:summation_density_resets_done}    
\end{equation}
which can be readily inverted to have the desired equation for $\rho(t)$:
\begin{equation}
\rho(t)=q(t) \rho_{\mathrm{free},1}(t)+\int_{0}^{t} \mbox{d}\tau q(t-\tau) \rho_{\mathrm{free},1}(t-\tau)r_{11}(\tau) + 
\int_{0}^{t} \mbox{d}\tau q(t-\tau) \rho_{\mathrm{free},2}(t-\tau)r_{12}(\tau).
\label{eq:renewal_equation}
\end{equation}
The previous equation can be solved to get $\rho(t)$ once $r_{11}(t),r_{12}(t) \geq 0$ are known from the inverse Laplace transform of Eq.~\eqref{eq:rates}. Note that $r_{11}(t),r_{12}(t)$ are non-negative (and the same for $r_{21}(t)$ and $r_{22}(t)$) since they are obtained from the inverse Laplace transform of the series in Eq.~\eqref{eq:sum_geometric_matrix}, with the matrix $R_{jk}(\tau) \geq 0$ from the definition in Eq.~\eqref{eq:R_matrix}. These two coefficients can be therefore considered as the effective time-dependent rates of jumping into the reset state $1$ or $2$, respectively, at time $t$ having started at time $0$ from the state $\ket{1}$. Equation \eqref{eq:renewal_equation} can be seen as the equivalent of the so-called ``last renewal equation'' of reset process, see, e.g., Refs.~\onlinecite{evans2011resetting,evans2011resettinglong,evans2020review} for classical systems and Refs.~\onlinecite{Mukherjee2018} for quantum systems. The fundamental complication with respect to the case of state-independent resets analyzed in \cite{Mukherjee2018} is encoded into the rates $r_{11}(t),r_{12}(t)$ which depend on the conditional reset mechanism, in our case see Eq.~\eqref{eq:probability_two_resets}. 

Note that the first term on the right hand side of Eq.~\eqref{eq:renewal_equation}, as well as the rates $r_{11}(t)$ and $r_{12}(t)$, depends on the choice of the initial state that we have done in Eq.~\eqref{eq:initial_state}. The stationary state, on the contrary, does not depend on this choice. The steady state limit of Eq.~\eqref{eq:renewal_equation} can be obtained from the ``final-value theorem'' of Laplace transforms \cite{feller1957introduction}, which for an arbitrary function of time $f(t)$ reads
\begin{equation}
\lim_{t \rightarrow \infty} f(t) = \lim_{s \rightarrow 0} s \widehat{f}(s), 
\label{eq:final_value_theorem}
\end{equation}
which gives the steady state limit by locating the singularity of $\widehat{f}(s)$ at $s=0$ (which must be a simple pole for the steady state to exist). From Eq.~\eqref{eq:matrix_R_norm}, one immediately sees that for $s=0$ one eigenvalue of $\widehat{R}(s=0)$ is equal to $1$, which causes the series in Eq.~\eqref{eq:sum_geometric_matrix} to diverge. The rates $\widehat{r}_{11}(s)$ and $\widehat{r}_{12}(s)$ (and the same for $\widehat{r}_{21}(s)$ and $\widehat{r}_{22}(s)$ which are not reported for the sake of brevity) in Eq.~\eqref{eq:rates} accordingly become singular at $s=0$ with a leading simple pole behavior behavior
\begin{align}
\widehat{r}_{11}(s) = \frac{\widehat{R}_{21}(0)}{s \, \,\widehat{q}(0)(\widehat{R}_{21}(0)+\widehat{R}_{12}(0))} \,\,\, \mbox{as} \, \, \, s \rightarrow 0 , \qquad \widehat{r}_{12}(s) = \frac{\widehat{R}_{12}(0)}{s \, \, \widehat{q}(0)(\widehat{R}_{21}(0)+\widehat{R}_{12}(0))}  \,\,\, \mbox{as} \, \, \, s \rightarrow 0,
\label{eq:rates_steady_expansion}
\end{align}
which inserted into Eq.~\eqref{eq:summation_density_resets} with Eq.~\eqref{eq:final_value_theorem} gives the steady state density matrix $\rho_{\mathrm{ness}}$
\begin{equation}
\rho_{\mathrm{ness}}=\lim_{t \rightarrow \infty} \rho(t) = \frac{\widehat{R}_{21}(0)}{\widehat{R}_{21}(0)+\widehat{R}_{12}(0)}\frac{1}{\widehat{q}(0)}\int_{0}^{\infty}\mbox{d}\tau \, \rho_{\mathrm{free},1}(\tau) q(\tau) + \frac{\widehat{R}_{12}(0)}{\widehat{R}_{21}(0)+\widehat{R}_{12}(0)}\frac{1}{\widehat{q}(0)}\int_{0}^{\infty}\mbox{d}\tau \, \rho_{\mathrm{free},2}(\tau) q(\tau).    
\label{eq:NESS_reset_1}
\end{equation}
This coincides with Eq.~(6) of the main text and it is one of the main results of this manuscript. We emphasize that the steady state $\rho_{\mathrm{ness}}$ is determined solely by the leading singular behavior of the rates around $s=0$ in Eq.~\eqref{eq:rates_steady_expansion}, which is independent on the choice of the initial state in Eq.~\eqref{eq:initial_state}. From Eq.~\eqref{eq:NESS_reset_1} one has access to the steady state average of an arbitrary observable $\mathcal{O}$ in terms of its free-unitary dynamics as
\begin{equation}
\Braket{\mathcal{O}}_{\mathrm{ness}}= \frac{\widehat{R}_{21}(0)}{\widehat{R}_{21}(0)+\widehat{R}_{12}(0)}\frac{1}{\widehat{q}(0)}\int_{0}^{\infty}\mbox{d}t \, \Braket{\mathcal{O}}_{\mathrm{free},1} q(t) + \frac{\widehat{R}_{12}(0)}{\widehat{R}_{21}(0)+\widehat{R}_{12}(0)}\frac{1}{\widehat{q}(0)}\int_{0}^{\infty}\mbox{d}t \, \Braket{\mathcal{O}}_{\mathrm{free},2} q(t), \label{eq:NESS_reset_2}    
\end{equation}
with the notation from Eqs.~\eqref{eq:unitary_evolution_states} and \eqref{eq:unitary_evolution_density_matrix}
\begin{equation}
\Braket{\mathcal{O}}_{\mathrm{free},j} = \Braket{\psi_j(t)|\mathcal{O}|\psi_j(t)}, \, \, \, \mbox{with} \, \, j=1,2.    
\end{equation}
In concluding this Subsection we mention that the weights of the terms in the sum in the right hand side of Eq.~\eqref{eq:NESS_reset_1} have a clear interpretation in terms of the underlying semi-Markov process. As a matter of fact, one immediately realizes that the vector 
\begin{equation}
\mathcal{P_{\mathrm{stat}}}=
\begin{pmatrix}
\widehat{R}_{21}(0)/ (\widehat{R}_{21}(0)+\widehat{R}_{12}(0)) \\
\widehat{R}_{12}(0)/ (\widehat{R}_{21}(0)+\widehat{R}_{12}(0))
\end{pmatrix}    
\label{eq:steady_rates}
\end{equation}
is a left eigenvector of $\widehat{R}(s=0)$ with eigenvalue $1$. The entries $\mathcal{P}_{\mathrm{stat}}^{(1)}$ and $\mathcal{P}_{\mathrm{stat}}^{(2)}$ of the vector $\mathcal{P}_{\mathrm{stat}}$ are nothing but the steady states occupation probabilities of a two-state Markov chain \cite{berman1994nonnegative}. On the basis of this observation it becomes also evident that the generalization of the derivation of $\rho_{\mathrm{ness}}$ to an arbitrary number $n$ of reset states, $\ket{1}$, $\ket{2}\dots \ket{n}$, is straightforward. Equation \eqref{eq:NESS_reset_1} generalizes to this case by embodying the summation over $n$ terms corresponding to the reset-free evolution starting from each of the $n$ reset states. The weight of each term of the sum will be then given by the steady state occupation probability of the associated $n$-state Markov chain identified by the $n \times n$ matrix $\widehat{R}_{jk}(s=0)$, $j,k=1,2\dots n$, in the Laplace $s$ domain.

\subsection{Generalized master equation approach and the Lindblad equation}
\label{sec:ness_master_eq}
We now show that Eq.~\eqref{eq:renewal_equation} for $\rho(t)$ can be seen as the solution of an evolution equation for $\mbox{d}\rho/\mbox{d}t$. To do this, we first define the operator $\mathcal{L}$:
\begin{equation}
\mathcal{L}[\rho] = -i[H,\rho], \quad \rho_{\mathrm{free},i}(t) = \mbox{exp}(\mathcal{L}t)\ket{i}\bra{i}, \label{eq:unitary_generator}    
\end{equation}
which is the generator of the reset-free unitary time evolution $\rho_{\mathrm{free},i}(t)$ of the density matrix. One, then, proceeds from Eq.~\eqref{eq:summation_density_resets_done}, which we rewrite as 
\begin{equation}
\widehat{\rho}(s) = \widehat{q}(s-\mathcal{L})\ket{1}\bra{1} + \widehat{r}_{11}(s)  \widehat{q}(s-\mathcal{L})  \ket{1}\bra{1} + \widehat{r}_{12}(s) \widehat{q}(s-\mathcal{L})  \ket{2}\bra{2}.
\label{eq:intermediate_master_equation}
\end{equation}
From the relation between the survival probability $q(\tau)$ and the waiting time probability density $p(\tau)$ in Eq.~\eqref{eq:survival_probability} in the Laplace-$s$ domain 
\begin{equation}
\widehat{q}(s) = \frac{1-\widehat{p}(s)}{s},
\label{eq:laplace_survival}
\end{equation}
and the shift property of the Laplace transform applied to Eq.~\eqref{eq:rho_up_down_def}, one recasts Eq.~\eqref{eq:intermediate_master_equation} as 
\begin{equation}
s \widehat{\rho}(s)-\ket{1}\bra{1} = \mathcal{L}[\widehat{\rho}(s)] +\widehat{r}_{11}(s) \ket{1}\bra{1}+\widehat{r}_{12}(s)\ket{2}\bra{2}-\widehat{r}(s-\mathcal{L})\widehat{\rho}(s),
\label{eq:intermediate_2_master_equation}
\end{equation}
with 
\begin{equation}
\widehat{r}(s) = \int_{0}^{\infty}\mbox{d}\tau \, r(\tau) e^{-s \tau}= \frac{s \, \widehat{p}(s)}{1-\widehat{p}(s)}=\frac{ \widehat{p}(s)}{\widehat{q}(s)}.
\label{eq:total_rate}
\end{equation}
Since $\rho(0)=\ket{1}\bra{1}$ from Eq.~\eqref{eq:initial_state}, one can invert Eq.~\eqref{eq:intermediate_2_master_equation} to the time domain eventually obtaining
\begin{equation}
\frac{\mbox{d}\rho(t)}{\mbox{d}t} = \mathcal{L}[\rho(t)]+r_{11}(t)\ket{1}\bra{1}+r_{12}(t)\ket{2}\bra{2}-\int_{0}^{t} \mbox{d}t' r(t-t') e^{(t-t')\mathcal{L}} \rho(t').
\label{eq:generalized_master_equation}
\end{equation}
Note that the rates $r_{11}(t),r_{12}(t)$ have to be related to $r(t)$ in order for Eq.~\eqref{eq:generalized_master_equation} to be trace-preserving, in particular, one needs the constraint
\begin{equation}
r_{11}(t)+r_{12}(t) = \int_{0}^{t} \mbox{d}t' r(t'),
\label{eq:rates_equation}
\end{equation}
which can be readily checked in Laplace space as 
\begin{equation}
\widehat{r}_{11}(s)+\widehat{r}_{12}(s) = \frac{\widehat{r}(s)}{s} = \frac{\widehat{p}(s)}{1-\widehat{p}(s)},    
\end{equation}
with simple algebraic manipulations starting from Eq.~\eqref{eq:rates}. 
The first term on the right hand side of equation \eqref{eq:generalized_master_equation} describes the unitary dynamics between reset events according to Eq.~\eqref{eq:unitary_generator}. The second and third terms, containing the effective rates $r_{11}(t)$ and $r_{12}(t)$, can be seen as gain terms due to resetting trajectories jumping into the states $\ket{1}$ and $\ket{2}$. The last term on the right hand side, on the other hand, can be considered as a loss term due to resetting trajectories where the systems evolves unitarily in the time interval $t-t'$ and it is then subject to resetting. Equation \eqref{eq:generalized_master_equation} can be therefore considered as a  generalized master equation. The derivation here adopted extends the one of Ref.~\onlinecite{eule2016non}, which regards state-independent reset processes, by including the effect of the conditional reset protocol within the rates $r_{11}(t),r_{12}(t)$ (analogously to the discussion done after Eq.~\eqref{eq:renewal_equation}). The latter can be written in terms of the density matrix as 
\begin{equation}
r_{1k}(t) = \int_{0}^{t}\mbox{d}t' r(t-t')\mbox{Tr}\left[\mathcal{P}_k e^{(t-t')\mathcal{L}}\rho(t')\right], \quad \mbox{with} \quad k=1,2,
\label{eq:non_Markov_rates_suppl}    
\end{equation}
and the operator $\mathcal{P}_k$ defined in Eq.~\eqref{eq:projector}. Equation \eqref{eq:non_Markov_rates_suppl} can be proved by replacing $\rho(t')$ inside the integral with Eq.~\eqref{eq:renewal_equation} and integrating each term. In particular, the integration of the first term on the right hand side of Eq.~\eqref{eq:renewal_equation} gives
\begin{equation}
\int_{0}^{t}\mbox{d}t' r(t-t')\mbox{Tr}\left[\mathcal{P}_k e^{(t-t')\mathcal{L}}q(t') \rho_{\mathrm{free},1}(t')\right] = \int_{0}^{t}\mbox{d}t' r(t-t')q(t') \mbox{Tr}\left[\mathcal{P}_k  \rho_{\mathrm{free},1}(t)\right] = p(t) P_{1k}(t)= R_{1k}(t),
\label{eq:step_1_non_markov_rates}
\end{equation}
where in the second equality we used the definitions in Eqs.~\eqref{eq:unitary_evolution_density_matrix} and \eqref{eq:unitary_generator} for the reset-free, unitary, dynamics, and Eq.~\eqref{eq:total_rate}. The integral of the second and third term of Eq.~\eqref{eq:renewal_equation} can be similarly performed 
\begin{align}
\int_{0}^{t}\mbox{d}t' \int_{0}^{t'}\mbox{d}\tau \, r(t-t')\mbox{Tr}\left[\mathcal{P}_k e^{(t-t')\mathcal{L}}q(t'-\tau)\rho_{\mathrm{free},j}(t'-\tau)r_{1j}(\tau)\right] &= \int_{0}^t \mbox{d}\tau P_{jk}(t-\tau)r_{1j}(\tau)\int_{\tau}^t \mbox{d}t' r(t-t')q(t'-\tau)\nonumber \\
&=\int_{0}^t\mbox{d}\tau \, r_{1j}(\tau) R_{jk}(t-\tau),
\label{eq:step_2_non_markov_rates}
\end{align}
where in the second equality we exchanged the order of integration. Inserting Eqs.~\eqref{eq:step_1_non_markov_rates} and \eqref{eq:step_2_non_markov_rates} into Eq.~\eqref{eq:non_Markov_rates_suppl} one has \begin{equation}
r_{1k}(t)= R_{1k}(t)+\sum_{j=1}^2 \int_{0}^t \mbox{d}\tau \, r_{1j}(\tau) R_{jk}(t-\tau),  \quad \mbox{with} \, \, \, k=1,2.
\label{eq:step_3_non_markov_rates}
\end{equation}
The vector-matrix multiplication on the r.h.s. of Eq.~\eqref{eq:step_3_non_markov_rates} can be performed using Eq.~\eqref{eq:rates}. Upon taking the Laplace transform of the convolution integral on the r.h.s. of Eq.~\eqref{eq:step_3_non_markov_rates} one, indeed, has
\begin{equation}
\sum_{j=1}^2 \widehat{r}_{1j}(s) \widehat{R}_{jk}(s) = \sum_{j=1}^2 ((\mathbb{I}-\widehat{R}(s))^{-1}-\mathbb{I})_{1j} R_{jk}(s) = ((\mathbb{I}-\widehat{R}(s))^{-1}-\mathbb{I})_{1k}-\widehat{R}(s)_{1k}=\widehat{r}_{1k}(s)-\widehat{R}_{1k}(s), \quad \mbox{with} \, \, \, k=1,2,
\label{eq:matrix_vector_rates_final}
\end{equation}
where in the last step we used the identity
\begin{align}
\sum_{j=1}^2 ((\mathbb{I}-\widehat{R}(s))^{-1})_{1j} \widehat{R}_{jk}(s) = \sum_{j=1}^2 \sum_{n=0}^{\infty} (\widehat{R}^n(s))_{1j}\widehat{R}(s)_{jk} = \sum_{n=0}^{\infty} \sum_{j=1}^2 (\widehat{R}^n(s))_{1j}\widehat{R}(s)_{jk} &= \sum_{n=0}^{\infty} (\widehat{R}^{n+1}(s))_{1k}  \nonumber \\ 
&=((\mathbb{I}-\widehat{R}(s))^{-1}-\mathbb{I})_{1k}.
\label{eq:matrix_vector_rates_intermediate_2}
\end{align}
Upon inserting Eq.~\eqref{eq:matrix_vector_rates_final} into Eq.~\eqref{eq:step_3_non_markov_rates} (in Laplace domain), Eq.~\eqref{eq:non_Markov_rates_suppl} is eventually proved. 

Crucially Eq.~\eqref{eq:generalized_master_equation}, with the time dependent rates $r_{1k}(t)$ in Eq.~\eqref{eq:non_Markov_rates_suppl}, is non-Markovian since it contains a time-integral over the entire previous evolution of the process. We can however rewrite Eq.~\eqref{eq:generalized_master_equation} with Eq.~\eqref{eq:non_Markov_rates_suppl} in the form of a generalized Lindblad equation 
\begin{align}
\frac{\mbox{d}\rho(t)}{\mbox{d}t} = \mathcal{L}[\rho(t)]&+\int_{0}^{t} \mbox{d}t' r(t-t')\left( \sum_{a=0_{1},1}\sum_{\mu} L_{\mu}^a \left[e^{(t-t')\mathcal{L}}\rho(t') \right](L_{\mu}^a)^{\dagger} +\sum_{a=0_{2},2}\sum_{\mu} L_{\mu}^a \left[e^{(t-t')\mathcal{L}} \rho(t') \right](L_{\mu}^a)^{\dagger} \right) \nonumber \\
&-\int_{0}^{t} \mbox{d}t' r(t-t') \sum_{a=0_1,0_2,1,2}\sum_{\mu} \frac{1}{2}\left\{(L_{\mu}^{a})^{\dagger} L_{\mu}^{a}, e^{(t-t')\mathcal{L}} \rho(t')\right\}.
\label{eq:generalized_master_equation_jump_operators}
\end{align}
by defining the following sets of $N_L=2^N+\frac{N!}{((N/2)!)^2}$ jump operators 
\begin{subequations}
\begin{align}
L^{0_{1}} &= \left\{L^{0_1}_{\mu} = \sqrt{\frac{1}{2}} \ket{1}\bra{C_{\mu}^0}, \quad \mu=1,2 \dots \, \, \frac{N!}{((N/2)!)^2} \right\},   \\ 
L^{0_{2}} &= \left\{L^{0_2}_{\mu} = \sqrt{\frac{1}{2}} \ket{2}\bra{C_{\mu}^0}, \quad \mu=1,2 \dots \, \, \frac{N!}{((N/2)!)^2} \right\}, \\
L^1 &= \left\{L^1_{\mu} =  \ket{1}\bra{C_{\mu}^1}, \quad \mu=1,2\dots \, \, 2^{N-1}-\frac{N!}{2 ((N/2)!)^2}  \right\}, \\
L^2 &= \left\{L^2_{\mu} = \ket{2}\bra{C_{\mu}^2}, \quad \mu=1,2 \dots 2^{N-1}-\frac{N!}{2 ((N/2)!)^2}  \right\}.
\end{align}
\label{eq:reset_jump_operators_exp}%
\end{subequations}
In Eq.~\eqref{eq:generalized_master_equation_jump_operators} each summation on the index $\mu$ runs over the range specified by Eq.~\eqref{eq:reset_jump_operators_exp} and $\{\dots,\dots \}$ denotes the anticommutator. Notice that in Eq.~\eqref{eq:generalized_master_equation_jump_operators} we used the definition of $\mathcal{P}_k$ in Eq.~\eqref{eq:projector} and the fact that 
\begin{equation}
\sum_{a=0_{1},0_2,1,2} \sum_{\mu}\left(L_{\mu}^{a}\right)^{\dagger} L_{\mu}^{a} = \sum_{a=0,1,2} \sum_{\mu} \ket{C_{\mu}^a}\bra{C_{\mu}^a}= \mathbb{I}.
\label{eq:completeness_jump_operators}
\end{equation}

The case of Poissonian resetting at constant (time-independent) rate $\gamma$ is in this perspective special. In this case, on the basis of the definition of a Poisson process, a reset occurs in the time interval $(\tau,\tau +\mbox{d}\tau)$ with probability $\gamma \mbox{d}\tau$, without the need to know when the previous reset occurred. Particularizing Eq.~\eqref{eq:total_rate} to the exponential waiting-time distribution $p_{\gamma}(\tau)=\gamma \mbox{exp}(-\gamma \tau)$ one, indeed, has 
\begin{equation}
\widehat{r}(s) = \gamma, \quad r(t) =\gamma \, \delta(t),
\label{eq:total_rate_exp_case}
\end{equation}
and therefore Eq.~\eqref{eq:generalized_master_equation_jump_operators} simplifies to 
\begin{equation}
\frac{\mbox{d}\rho(t)}{\mbox{d}t} = \mathcal{L}[\rho(t)]+ \gamma\sum_{a=0_{1},1}\sum_{\mu} L_{\mu}^a \, \rho(t) (L_{\mu}^a)^{\dagger} +\gamma\sum_{a=0_{2},2}\sum_{\mu} L_{\mu}^a \, \rho(t) (L_{\mu}^a)^{\dagger}
-\gamma\sum_{a=0_1,0_2,1,2}\sum_{\mu} \frac{1}{2}\left\{(L_{\mu}^{a})^{\dagger} L_{\mu}^{a},\rho(t)\right\},
\label{eq:state_dependent_exp_master_equation}
\end{equation}
which is in the Lindblad form \cite{lindblad1976generators,gorini1976completely} 
\begin{equation}
\frac{\mbox{d} \rho(t)}{\mbox{d} t} = \mathcal{L}[\rho(t)] +
\sum_{a=0_1,0_2,1,2}\sum_{\mu}\left(L_{\mu}^a \rho(t) (L_{\mu}^a)^{\dagger} -\frac{1}{2} \{(L_{\mu}^a)^{\dagger} L_{\mu}^a, \rho(t)\} \right),
\label{eq:Lindblad}
\end{equation}
with the jump operators $\sqrt{\gamma}L_{\mu}^a \to L_{\mu}^a$ rescaled by the resetting rate $\gamma$. Equation \eqref{eq:state_dependent_exp_master_equation} is, fundamentally, Markovian since it is local in time, differently from Eq.~\eqref{eq:generalized_master_equation_jump_operators}. 
This derivation extends the analysis of Ref.~\onlinecite{Hartmann2006,Linden2010,Armin2020,Rose2018spectral}, where only state-independent (no conditional reset mechanism) reset was considered. Fundamentally, in Ref.~\onlinecite{Hartmann2006,Linden2010,Armin2020,Rose2018spectral} the Lindblad equation is assumed as a starting point for the analysis of Poissonian resetting at rate $\gamma$. The analysis of this manuscript, on the contrary, shows exactly how the averaging of the microscopic unitary evolution over the reset trajectories leads to an effective non-Markovian generalized Lindblad equation \eqref{eq:generalized_master_equation_jump_operators}. In the specific case of the exponential waiting time probability density $p_{\gamma}(\tau)$, our analysis constitute an exact derivation of the Markovian Lindblad equation for an open quantum system subject to stochastic resetting.

We mention that for the exponential waiting-time distribution one can simulate the conditional reset dynamics with the stochastic wave function approach \cite{plenio1998quantumjumps}, where the probability $\mathrm{Prob}_{jk}$ of making a jump (reset) towards the state $\ket{k}$, being the last jump (reset) to $\ket{j}$, is
\begin{equation}
\mathrm{Prob}_{jk} = \Braket{\widetilde{\psi}_j(\tau)|\sum_{a=0_k,k}\sum_{\mu}\left(L_{\mu}^{a}\right)^{\dagger} L_{\mu}^{a}|\widetilde{\psi}_j(\tau)}= \gamma e^{-\gamma \tau} \Braket{\psi_j(\tau)| \mathcal{P}_k | \psi_j(\tau)}= P_{jk}(\tau) p_{\gamma}(\tau) =R_{jk}(\tau), \label{eq:stochastic_wave_function_algorithm}
\end{equation}
where
\begin{equation}
\ket{\widetilde{\psi}_j(\tau)} = \mbox{exp}(-i H_{\mathrm{eff}}\tau)\ket{j}, \, \, \, \quad H_{\mathrm{eff}}= H-\frac{i}{2} \sum_{a=0_{1},0_{2},1,2} \sum_{\mu}\left(L_{\mu}^{a}\right)^{\dagger} L_{\mu}^{a} = H -\frac{i\gamma}{2}, 
\end{equation}
with $H_{\mathrm{eff}}$ denoting the effective, non-Hermitean, Hamiltonian which dictates the time evolution $\ket{\widetilde{\psi}_j(\tau)}$ between consecutive quantum jumps (resets) \cite{plenio1998quantumjumps} and we have used the definitions in Eqs.~\eqref{eq:probability_two_resets} and Eq.~\eqref{eq:R_matrix}.  

\section{Stochastic resetting in the quantum Ising chain in a transverse field}
\label{sec:TFIC}
In this Section we particularize the general theory of Sec.~\ref{sec:ness} to the TFIC. Our motivation to investigate the TFIC model is twofold: first, it is a paradigmatic model of (quantum) statistical mechanics which displays a quantum phase transition; second, it allows for an exact determination of the NESS with reset in the thermodynamic limit. The availability of such an analytical result is particularly useful as a reference for the analysis of more complex models, such as Rydberg atoms arrays \cite{RydbergReview2020}, where a full analytical treatment is not possible. In Subsec.~\ref{sec:equilibrium_TFIC} we briefly recapitulate some well known equilibrium properties of the quantum Ising chain which will be essential for the understanding of our results. In Subsec.~\ref{sec:resetting_TFIC} we discuss the dynamics of the TFIC for the conditional reset protocol of Sec.~\ref{sec:ness}. In Subsec.~\ref{sec:phase_diagram_TFIC} we give details of the evaluation of the phase diagram for $m^2_{\mathrm{ness}}$, which is reported in Fig.~2 of the main text. In Subsec.~\ref{sec:TFIC_asympt} we eventually detail the derivation of the asymptotics of $m^2_{\mathrm{ness}}$ for large $J t_{\mathrm{max}}$, low $\gamma/J$, which is reported in Fig.~2(d) of the main text.

\subsection{Equilibrium properties of the Ising chain in a transverse field}
\label{sec:equilibrium_TFIC}
The Hamiltonian of the model for a total number $N$ of spins is given by
\begin{equation}
H= -J \sum_{n=1}^{N} (\sigma_n^x \sigma_{n+1}^x +h\sigma_n^z),
\label{eq:TFIC_Hamiltonian}
\end{equation}
where we assume periodic boundary conditions $\sigma^x_{n+N}=\sigma^x_{n}$. Various notational conventions can be adopted for the TFIC, here we follow the one of Refs.~\onlinecite{calabrese2011quantum,calabrese2012quantum1,calabrese2012quantum2}. The ferromagnetic coupling $J>0$ gives an overall energy scale. Time is accordingly measured in units of $J^{-1}$. The quantum Ising chain is a paradigmatic example of a quantum phase transition happening for the critical magnetic field value $h_c=1$ at zero temperature $T=0$. The paramagnetic-disordered phase is attained for $h>1$, while the ferromagnetic-ordered phase for $h<1$. In the latter phase the order parameter
\be
m=\frac{1}{N}\sum_{n=1}^N \sigma^x_n
\label{eq:magnetization_density}
\ee
of the quantum phase transition becomes non-zero. The quantum phase transition comes with the spontaneous breaking of the $\mathbb{Z}_2$ symmetry of rotations by $180$° around the $z$ axis 
\begin{equation}
\sigma_j^{\alpha} \rightarrow -\sigma_j^{\alpha}, \quad \mbox{with} \, \, \alpha=x,y, \quad \sigma_j^{z} \rightarrow \sigma_j^z.
\label{eq:Z_2_rotations}
\end{equation}
The quantum Ising Hamiltonian in Eq.~\eqref{eq:TFIC_Hamiltonian} possesses the $\mathbb{Z}_2$ symmetry, which is implemented by the unitary operator $U$:
\begin{equation}
\mathcal{U} = \prod_{j=1}^N \sigma_j^z, \quad \mathcal{U} H \mathcal{U}^{\dagger} =H.
\label{eq:Z_2_symmetry_Hamiltonian}
\end{equation}
We define lattice fermion operators using the Jordan-Wigner transformation (see, e.g., Refs.~\onlinecite{lieb1961two,kogut1979})
\begin{equation}
c_n^{\dagger} = \left(\prod_{j=1}^{n-1} \sigma_j^z\right) \sigma_n^-, \quad \sigma_n^{\pm} = \frac{\sigma_n^x \pm i\sigma_n^y}{2}, \quad \sigma_n^z = 1- 2c_n^{\dagger} c_n,    
\label{eq:JW_transformation}
\end{equation}
in terms of which the operator $\mathcal{U}$ in Eq.~\eqref{eq:Z_2_symmetry_Hamiltonian} can be written as
\begin{equation}
\mathcal{U}= e^{i \pi \mathcal{N}}, \quad \mathcal{N}=\sum_{n=1}^N c_n^{\dagger} c_n.
\end{equation}
The unitary operator $\mathcal{U}$ counts, therefore, the parity of the Jordan-Wigner fermions $c_n$. As a consequence of the $\mathbb{Z}_2$ symmetry, $H$ and $\mathcal{U}$ can be diagonalized simultaneously and the Hamiltonian is block diagonal 
\be
H=H_e \frac{\mathbb{I}+\mathcal{U}}{2} + H_o \frac{\mathbb{I}-\mathcal{U}}{2},
\label{eq:even_odd_Hamiltonian_decomposition}
\ee
where $H_e$ and $H_o$ denote the Hamiltonian restricted to the even and odd fermion sector of the Hilbert space, respectively. The exact solution of the model in terms of free fermionic quasi-particles is extremely well known, see, e.g., Refs.~\onlinecite{lieb1961two,kogut1979}, and we therefore do not report it here for the sake of brevity. It is sufficient for our purpose to report the single-quasi-particle energy spectrum $\varepsilon_h(k)$ 
\begin{equation}
\varepsilon_h(k) = 2J \sqrt{h^2-2h \, \cos k+1}.
\label{eq:TFIC_spectrum}
\end{equation}
The ground state of $H_e$ in the even fermion sector will be denoted henceforth as $\ket{0,h}_{NS}$ (the subscript NS stands for Neveu-Schwarz, which is the name conventionally given to the even fermion sector \cite{calabrese2011quantum,calabrese2012quantum1,calabrese2012quantum2}). The ground state of $H_o$ in the odd fermion sector will be denoted as $\ket{0,h}_R$ (the subscript R stands for Ramond which is the name given to the odd fermion sector). 
Note that 
\begin{equation}
\mathcal{U} \ket{0,h}_{NS} = \ket{0,h}_{NS}, \quad \mathcal{U} \ket{0,h}_{R} = \mbox{sgn}(h-1) \ket{0,h}_R, 
\label{eq:GS_symmetry}
\end{equation}
which shows that the NS and R vacua are eigenstates of the parity operator $\mathcal{U}$, i.e., they do not break the $\mathbb{Z}_2$ symmetry. For $h>1$, in the paramagnetic phase, it can be shown \cite{calabrese2011quantum,calabrese2012quantum1,calabrese2012quantum2} that $\ket{0,h}_{NS}$ is the ground state of the complete Hamiltonian $H$ in Eq.~\eqref{eq:even_odd_Hamiltonian_decomposition} since it has an energy lower than the one of $\ket{0,h}_{R}$. For $h<1$, in the ferromagnetic phase, the spontaneous breaking of the $\mathbb{Z}_2$ symmetry is reflected by the fact that, in the thermodynamic limit $N \rightarrow \infty$, the states $\ket{0,h}_{NS}$ and $\ket{0,h}_R$ become degenerate. As a consequence in the thermodynamic limit, for $h<1$, either the symmetric $\ket{\mbox{GS}(h)}_{+}$ or the anti-symmetric $\ket{\mbox{GS}(h)}_{-}$ combination of $\ket{0,h}_{NS}$ and $\ket{0,h}_R$ is chosen as a ground state.
In formulas, the ground state of the system will be denoted as $\ket{\mbox{GS}(h)}$ and it is then given by 
\be
\ket{\mbox{GS}(h)} = \left\{
  \begin{array}{lr}
    \ket{\mbox{GS}(h)}_{\pm}=\frac{\ket{0,h}_{NS}\pm \ket{0,h}_{R}}{\sqrt{2}} \; \; \; \;  \mbox{for} \; \; \; h<1; \\
    \ket{0,h}_{NS} \qquad \qquad \qquad \qquad  \; \; \, \mbox{for} \; \; \; \;  h>1.
  \end{array}
\right. \label{eq:GS_TFIC}
\ee
Note that from Eq.~\eqref{eq:GS_symmetry}
\be
\mathcal{U} \ket{\mbox{GS}(h)}_{\pm}=\mathcal{U}\left(\frac{\ket{0,h}_{NS}\pm \ket{0,h}_{R}}{\sqrt{2}}\right) = \frac{\ket{0,h}_{NS}\mp \ket{0,h}_{R}}{\sqrt{2}} = \ket{\mbox{GS}(h)}_{\mp}.
\label{eq:GS_symmetry_broken}
\ee
Namely, the ground state for $h<1$ is not an eigenstate of the fermion parity, thereby expressing the breaking of the $\mathbb{Z}_2$ symmetry. Spontaneous symmetry breaking is also detected in the ground-state expectation value of the order parameter $m$ in Eq.~\eqref{eq:magnetization_density}, which is non-vanishing in the symmetry-broken phase for $h<1$ \cite{lieb1961two,McCoy1,McCoy2,McCoy3,McCoy4}
\begin{equation}
\tensor*[_{\pm}]{\Braket{\mbox{GS}(h)|m|\mbox{GS}(h)}}{_{\pm}} = \tensor*[_{\pm}]{\Braket{\mbox{GS}(h)|\sigma^x|\mbox{GS}(h)}}{_{\pm}} =\pm (1-h^2)^{1/8}.
\label{eq:GS_magnetization}
\end{equation}
In the first equality of the previous equation we used the fact that for periodic boundary conditions the chain is translationally invariant and therefore the vacuum expectation value  ${}_{\pm}\Braket{\mbox{GS}(h)|\sigma^x_n|\mbox{GS}(h)}_{\pm}$ does not depend on the lattice site $n$. We conclude this subsection by defining even $\mathcal{O}_e$ and odd $\mathcal{O}_o$ operators under the $\mathbb{Z}_2$ transformation in Eq.~\eqref{eq:Z_2_rotations} as 
\begin{equation}
\mathcal{U} \left(\mathcal{O}_{e/o}\right) \mathcal{U^{\dagger}} = \pm \mathcal{O}_{e/o}, \label{eq:even_odd_operators}   
\end{equation}
with the plus sign applying to even operators, while the minus sign to odd operators. The distinction between even and odd operators is crucial for the study of the dynamics of the TFIC from the reset states $\ket{1}$ and $\ket{2}$ in Eq.~\eqref{eq:reset_states} according to the formalism explained in Sec.~\ref{sec:ness}. This is the problem that we address in the next Subsection.

\subsection{Construction of the NESS for the quantum Ising chain}
\label{sec:resetting_TFIC}
In this Subsection we construct the steady state density matrix $\rho_{\mathrm{ness}}$ and the corresponding stationary expectation values according to the general theory of Sec.~\ref{sec:ness} (see Eqs.~\eqref{eq:NESS_reset_1} and \eqref{eq:NESS_reset_2}). We do this by connecting our theory with quantum quenches in the Ising chain.
We first note that the choice of the reset states $\ket{1}$ and $\ket{2}$ in Eq.~\eqref{eq:reset_states} allows for a clear connection with quantum quenches since $\ket{1}=\ket{GS(0)}_{+}$ and $\ket{2}=\ket{GS(0)}_{-}$ are the two degenerate ground states of the Ising Hamiltonian $H(h_0)$ for $h_0=0$, according to Eq.~\eqref{eq:GS_symmetry_broken}. The unitary time evolution between two consecutive reset events is therefore a quench of the transverse field \cite{Sachdev2004quenches,quench2,quench4,quench5,calabrese2011quantum,calabrese2012quantum1,calabrese2012quantum2} from the pre-quench value $h_0=0$ to the post-quench value $h$. 
To make contact with the notation introduced in the previous Subsec.~\ref{sec:equilibrium_TFIC} we write the reset states as follows 
\begin{subequations}
\begin{align}
\ket{0,h_0=0}_{NS} &= \frac{1}{\sqrt{2}}\left(\ket{1} +\ket{2}\right), \quad \ket{1} =  \frac{1}{\sqrt{2}}\left(\ket{0,h_0=0}_{NS} + \ket{0,h_0=0}_{R} \right),  \\
\ket{0,h_0=0}_{R} &= \frac{1}{\sqrt{2}}\left(\ket{1} -\ket{2}\right), \quad \ket{2} =  \frac{1}{\sqrt{2}}\left(\ket{0,h_0=0}_{NS} - \ket{0,h_0=0}_{R} \right).
\end{align}
\label{eq:reset_states_Z_2_symmetry}%
\end{subequations}
We observe that, for $h_0=0$, the spin representation of the fermionic NS and R vacua is simple, as it corresponds to the superposition of the two eigenstates completely polarized in the $x$-longitudinal direction of the Hamiltonian in Eq.~\eqref{eq:TFIC_Hamiltonian}. For generic $h_0 \neq 0$ the representation of the NS and R vacua in the spin language cannot be simply written. This also motivates our choice for the reset states $\ket{1}=\ket{GS(0)}_{+}$ and $\ket{2}=\ket{GS(0)}_{-}$, since they have a simple representation in the spin basis as product states. This makes the experimental realization of such states simpler. We, however, observe that the following discussion can be immediately generalized , upon writing $\ket{1}=\ket{GS(h_0)}_{+}$ and $\ket{2}=\ket{GS(h_0)}_{-}$ in Eq.~\eqref{eq:reset_states_Z_2_symmetry}, to the case where the two reset states are taken as the two degenerate ground states for a non-zero value of the pre-quench transverse field $h_0$. For this reason, in all the following formulas, we denote with $h_0$ the pre-quench transverse field albeit in the numerical evaluation we have always used the specific value $h_0=0$.     
We now use Eq.~\eqref{eq:reset_states_Z_2_symmetry} to rewrite Eq.~\eqref{eq:NESS_reset_1} in a simpler way. First of all because of the $\mathbb{Z}_2$ symmetry, the expression we have chosen for the probability $P_{jk}(\tau)$ in Eq.~\eqref{eq:probability_two_resets} renders
\begin{equation}
P_{12}(\tau) = P_{21}(\tau), \quad P_{11}(\tau)=P_{22}(\tau),     
\label{eq:probability_symmetry}
\end{equation}
and therefore the matrix $\widehat{R}(s)$ in Eq.~\eqref{eq:R_matrix} has a very simple structure
\begin{equation}
\widehat{R}(s) = \mathbb{I} \, \widehat{R}_{11}(s) + \sigma^x \, \widehat{R}_{12}(s), \qquad \widehat{R}_{12}(s)=\widehat{R}_{21}(s), \qquad \widehat{R}_{11}(s)=\widehat{R}_{22}(s). \label{eq:R_matrix_Z_2}
\end{equation}
Using Eq.~\eqref{eq:R_matrix_Z_2} into Eq.~\eqref{eq:NESS_reset_1} one has
\begin{equation}
\rho_{\mathrm{ness}}=\lim_{t \rightarrow \infty} \rho(t) = \frac{1}{2}\frac{1}{\widehat{q}(0)}\int_{0}^{\infty}\mbox{d}\tau \, \rho_{\mathrm{free},1}(\tau) q(\tau) + \frac{1}{2}\frac{1}{\widehat{q}(0)}\int_{0}^{\infty}\mbox{d}\tau \, \rho_{\mathrm{free},2}(\tau) q(\tau),
\label{eq:NESS_reset_Z_2}
\end{equation}
where $\rho_{\mathrm{free,1}}(\tau)$ ($\rho_{\mathrm{free,2}}(\tau)$) is the quench-unitary time evolution over the time interval $\tau$ between two consecutive resets. It is important to emphasize that Eqs.~\eqref{eq:probability_symmetry}-\eqref{eq:NESS_reset_Z_2} are a consequence not only of the $\mathbb{Z}_2$ symmetry of the Hamiltonian \eqref{eq:TFIC_Hamiltonian}, but also of the chosen reset states in Eq.~\eqref{eq:reset_states} and of the form for the probability $P_{jk}(\tau)$ in Eq.~\eqref{eq:probability_two_resets}. 

We now want to draw some general statements regarding $\rho_{\mathrm{ness}}$ in Eq.~\eqref{eq:NESS_reset_Z_2} for the case of the waiting time probability density $p_{\gamma,t_{\mathrm{max}}}(\tau)$ (with the associated survival probability $q_{\gamma,t_{\mathrm{max}}}(\tau)$) in Eq.~\eqref{eq:chopped_exp}. In particular, the limiting case of the exponential distribution \eqref{eq:exponential_distribution} $t_{\mathrm{max}} \to \infty$ (with fixed $\gamma$) in Eq.~\eqref{eq:exponential_distribution} has been already considered in Ref.~\onlinecite{Mukherjee2018}, where it has been shown that $\rho_{\mathrm{ness}}$ retains non-vanishing off-diagonal matrix elements in the basis of the eigenstates of the post-quench Hamiltonian $H(h)$. We now show that the same applies in the more general case of the waiting time $p_{\gamma,t_{\mathrm{max}}}(\tau)$ for any finite value of $\gamma$ and/or $t_{\mathrm{max}}$. 
One can prove this by considering Eq.~\eqref{eq:NESS_reset_Z_2} for $\gamma=0$. In this limit the matrix elements of $\rho_{\mathrm{ness}}$ in the basis of the eigenstates of the post-quench Hamiltonian $H(h)$ are
\be
(\rho_{\mathrm{ness}})_{\alpha,\beta} = \frac{1}{2}\frac{1}{\widehat{q}(0)}\int_{0}^{t_{\mathrm{max}}}\mbox{d}\tau \, \, \frac{t_{\mathrm{max}}-\tau}{t_{\mathrm{max}}} \left((\rho_{\mathrm{free},1}(\tau))_{\alpha \beta} + (\rho_{\mathrm{free},2}(\tau))_{\alpha \beta}\right), \quad \mbox{with} \quad \widehat{q}(0) = \frac{t_{\mathrm{max}}}{2},
\label{eq:Diagonal_ensemble_1}
\ee
with 
\be
(\rho_{\mathrm{free},i})_{\alpha \beta}(\tau) = \Braket{\alpha|\rho_{\mathrm{free},i}(\tau)|\beta} =e^{i\omega_{\beta \alpha}\tau}c_{i\alpha}c_{i\beta}^{\ast}, \quad \omega_{\beta \alpha} = E_{\beta} -E_{\alpha}, \, \, i=1,2. 
\label{eq:Diagonal_ensemble_2}
\ee
In the previous equation $c_{\alpha} = \Braket{\alpha|i}$ denotes the overlap coefficients between the reset state $\ket{i}$ and the eigenstates $H(h)\ket{\alpha} = E_{\alpha}(h) \ket{\alpha}$ of the post-quench Hamiltonian. 
Inserting Eq.~\eqref{eq:Diagonal_ensemble_2} into Eq.~\eqref{eq:Diagonal_ensemble_1} and performing the integration one obtains
\begin{align}
(\rho_{\mathrm{ness}})_{\alpha,\beta} =\frac{2}{t_{\mathrm{max}}}\left(-\frac{1}{i \omega_{\beta \alpha}} + \frac{1-e^{i\omega_{\beta \alpha} t_{\mathrm{max}}}}{t_{\mathrm{max}}\omega_{\beta \alpha}^2} \right)\frac{1}{2}\left(c_{1\alpha}c_{1\beta}^{\ast}+c_{2\alpha}c_{2\beta}^{\ast} \right).
\label{eq:Diagonal_ensemble_3}
\end{align}
It is then clear that for any finite value of $t_{\mathrm{max}}$ the off-diagonal matrix elements of $\rho_{\mathrm{ness}}$ are different from zero. 
The steady state density matrix becomes diagonal in the basis of the post-quench eigenstates only as $t_{\mathrm{max}} \to \infty$
\be
(\rho_{\mathrm{ness}})_{\alpha \beta}= \left\{
  \begin{array}{lr}
    0 \; \; \; \;  \, \, \quad \qquad \qquad \qquad  \alpha \neq \beta, \\
     \frac{1}{2}\left(|c_1|^2+|c_2|^2\right) \quad \quad  \alpha=\beta.
  \end{array}
\right. \label{eq:Diagonal_ensemble_final}
\ee
Plugging Eq.~\eqref{eq:Diagonal_ensemble_final} into Eq.~\eqref{eq:NESS_reset_Z_2} one has
\be
\rho_{\mathrm{ness}}=\sum_{\alpha \beta} (\rho_{\mathrm{ness}})_{\alpha \beta}\ket{\alpha} \bra{\beta} = \sum_{\alpha} |c_1|^2 \ket{\alpha}\bra{\alpha},
\label{eq:Diagonal_ensemble_final_operator}
\ee
which coincides with the diagonal ensemble. In the last step we used that $|c_1|^2=|c_2|^2$ because of the $\mathbb{Z}_2$ symmetry. We therefore conclude that the steady state density matrix $\rho_{\mathrm{ness}}$ always retains non-vanishing off diagonal matrix elements in the basis of the eigenstates of the post-quench Hamiltonian $H(h)$ and it is therefore different from the diagonal ensemble, which describes the relaxation of closed quantum systems at long times \cite{ETH2016quantum}. Only in the case of $\gamma \to 0$ and $t_{\mathrm{max}} \to \infty$ the off-diagonal matrix elements vanish and the steady state density matrix in the presence of reset coincides with the diagonal ensemble, according to Eqs.~\eqref{eq:Diagonal_ensemble_final} and \eqref{eq:Diagonal_ensemble_final_operator}. This is expected because in the limit $\gamma \to 0$ and $t_{\mathrm{max}} \to \infty$, on average, a large time between consecutive resets elapses and therefore the dynamics is dominated by the long time asymptotics of the quench-unitary dynamics. The steady state density matrix $\rho_{\mathrm{ness}}$ is therefore  a non-equilibrium density matrix which is substantially different from the diagonal ensemble for any finite value of $\gamma$ and/or $t_{\mathrm{max}}$. 

The latter difference has clearly important consequences on the steady state expectation value of local observables. Inserting Eq.~\eqref{eq:NESS_reset_Z_2} into Eq.~\eqref{eq:NESS_reset_2} one has \be
\Braket{\mathcal{O}}_{\mathrm{ness}}= \mbox{Tr}(\mathcal{O} \rho_{\mathrm{ness}}) = \frac{1}{2}\frac{1}{\widehat{q}(0)}\int_{0}^{\infty}\mbox{d}\tau \, q(\tau) \left(\Braket{\mathcal{O}}_{\mathrm{free},1}(\tau) + \Braket{\mathcal{O}}_{\mathrm{free},2}(\tau)   \right).
\label{eq:NESS_Z2_1}
\ee
At this point, since the post-quench Hamiltonian $H(h)$ possesses the $\mathbb{Z}_2$ symmetry, it is important to distinguish between odd $\mathcal{O}_o$ and even $\mathcal{O}_e$ operators under $\mathbb{Z}_2$ according to Eq.~\eqref{eq:even_odd_operators}. We start from the case of odd operators, where using  Eq.~\eqref{eq:reset_states_Z_2_symmetry} one obtains 
\begin{equation}
\Braket{\mathcal{O}_o}_{\mathrm{free},1}(\tau) = -\Braket{\mathcal{O}_o}_{\mathrm{free},2}(\tau) = \mbox{Re}\left(\tensor*[_{NS}]{\Braket{0,h_0|\mathcal{O}_o(\tau)|0,h_0}}{_R}  \right), \quad \mbox{with} \quad \mathcal{O}_o(\tau) = e^{iH\tau} \mathcal{O}_o e^{-iH\tau}.    
\label{eq:odd_operators_Z2_symmetry}
\end{equation}
It is important to emphasize that the calculation of the expectation value of an odd operator over the symmetry-broken ground state is tremendously hard since it connects the ground state $\ket{0,h_0}_R$ in the odd fermion sector to the ground state $\ket{0,h_0}_{NS}$ in the even fermion sector, see, e.g., Refs.~\onlinecite{lieb1961two,McCoy1,McCoy2,McCoy3,McCoy4,calabrese2011quantum,calabrese2012quantum1,calabrese2012quantum2}. However, in the presence of resetting it follows by inserting Eq.~\eqref{eq:odd_operators_Z2_symmetry} into Eq.~\eqref{eq:NESS_Z2_1} that 
\be
\Braket{\mathcal{O}_o}_{\mathrm{ness}} \equiv 0 
\label{eq:odd_operators_zero_ness}
\ee
for any odd operator. For even operators, on the other hand, one has 
\begin{align}
\lim_{N \rightarrow \infty}\Braket{\mathcal{O}_e}_{\mathrm{free},1}(\tau) &= \lim_{N \rightarrow \infty} \Braket{\mathcal{O}_e}_{\mathrm{free},2}(\tau) = \lim_{N \rightarrow \infty} \frac{1}{2} (\tensor*[_{NS}]{\Braket{0,h_0|\mathcal{O}_e(\tau)|0,h_0}}{_{NS}} + \tensor*[_{R}]{\Braket{0,h_0|\mathcal{O}_e(\tau)|0,h_0}}{_{R}}) \nonumber \\ 
&=  \lim_{N \rightarrow \infty}\tensor*[_{NS}]{\Braket{0,h_0|\mathcal{O}_e(\tau)|0,h_0}}{_{NS}}, \quad \mbox{with} \quad \mathcal{O}_e(\tau) = e^{iH\tau} \mathcal{O}_e e^{-iH\tau},
\label{eq:even_operators_Z2_symmetry}
\end{align}
where we used the fact that averages of even operators $\mathcal{O}_e$ over the ground state $\ket{0,h_0}_{NS}$ are equal to the ones over the state $\ket{0,h_0}_R$ in the thermodynamic limit $N \to \infty$. By inserting in Eq.~\eqref{eq:even_operators_Z2_symmetry} into Eq.~\eqref{eq:NESS_Z2_1} one has
\be
\lim_{N \rightarrow \infty} \Braket{\mathcal{O}_e}_{\mathrm{ness}}=\frac{1}{\widehat{q}(0)}\int_{0}^{\infty}\mbox{d}\tau \, q(\tau) \lim_{N \rightarrow \infty} \tensor*[_{NS}]{\Braket{0,h_0|\mathcal{O}_e(\tau)|0,h_0}}{_{NS}}.
\label{eq:even_operators_ness}
\ee
The calculation of expectation values of even operators in the symmetry broken ground state according to Eq.~\eqref{eq:even_operators_Z2_symmetry} can be done by exploiting the Wick theorem-free fermions techniques. Exploiting the results available from Refs.~\onlinecite{lieb1961two,McCoy1,McCoy2,McCoy3,McCoy4,calabrese2011quantum,calabrese2012quantum1,calabrese2012quantum2} for the two-point correlation function of the order parameter $\sigma^x$ we derive in the next Subsection the exact result for the steady state squared magnetization density in the thermodynamic limit in the presence of resetting.

\subsection{Phase diagram for the magnetization density squared in the NESS}
\label{sec:phase_diagram_TFIC}
Based on Eq.~\eqref{eq:odd_operators_zero_ness} one has that the NESS expectation value $\Braket{m}_{\mathrm{ness}}$ is identically zero according to Eq.~\eqref{eq:odd_operators_zero_ness}, since the operator $m$ in Eq.~\eqref{eq:magnetization_density} is odd under the $\mathbb{Z}_2$ symmetry. To get a valuable order parameter distinguishing the ferromagnetic from the paramagnetic phase, the natural choice is then the squared magnetization density
\begin{equation}
m^2 = \frac{1}{N^2}\sum_{n=1}^N \sum_{m=1}^N \sigma^x_n \sigma^x_m,
\label{eq:magnetization_squared_order_parameter}
\end{equation}
which is even under the $\mathbb{Z}_2$ symmetry. Inserting Eq.~\eqref{eq:magnetization_squared_order_parameter} into Eq.~\eqref{eq:even_operators_ness} one can express everything in terms of the calculation of equal-time two-points correlation functions of the longitudinal magnetization  
\begin{align}
m^2_{\mathrm{ness}} \equiv \lim_{N \rightarrow \infty} \Braket{m^2}_{\mathrm{ness}}&=\frac{1}{\widehat{q}(0)}\int_{0}^{\infty}\mbox{d}\tau \, q(t) \lim_{N \rightarrow \infty} \frac{1}{N^2}\sum_{n=1}^{N}\sum_{m=1}^{N}\tensor*[_{NS}]{\Braket{0,h_0|\sigma^x_n(\tau) \sigma^x_m(\tau)|0,h_0}}{_{NS}} \nonumber \\
&= \frac{1}{\widehat{q}(0)}\int_{0}^{\infty}\mbox{d}\tau \, q(\tau) \lim_{N \rightarrow \infty} \frac{1}{N}\sum_{l=1}^{N}\tensor*[_{NS}]{\Braket{0,h_0|\sigma^x_n(\tau) \sigma^x_{n+l}(\tau)|0,h_0}}{_{NS}},
\label{eq:squared_m_intermediate}
\end{align}
where in the second line we have used translational invariance, which causes the two-point function to depend only on the distance $l$ between the two spins. We have also denoted, for brevity, $m^2_{\mathrm{ness}} = \lim_{N \to \infty} \Braket{m^2}_{\mathrm{ness}}$ to avoid writing henceforth the thermodynamic limit every time for the left hand side of Eq.~\eqref{eq:squared_m_intermediate}. One has to keep in mind that the following results apply in the thermodynamic limit $N \to \infty$. The summation on the right hand side of Eq.~\eqref{eq:squared_m_intermediate} is, indeed, dominated, as $N \to \infty$, by the thermodynamic limit of the equal-time two-point function of the order parameter 
\be
\rho^{xx}(l,t)=\lim_{N \rightarrow \infty} \tensor*[_{NS}]{\Braket{0,h_0|\sigma^x_n(t) \sigma^x_{n+l}(t)|0,h_0}}{_{NS}}. 
\label{eq:correlator_thermodynamic_limit}
\ee
Consequently, one can evaluate the sum in Eq.~\eqref{eq:squared_m_intermediate} directly replacing ${\tensor*[_{NS}]{\Braket{0,h_0|\sigma^x_n(t) \sigma^x_{n+l}(t)|0,h_0}}{_{NS}}}$ with $\rho^{xx}(l,t)$. The resulting summation can be evaluated using the Cesàro mean theorem \cite{borel}
\be
\lim_{N \rightarrow \infty}s_N=\lim_{N \rightarrow \infty} \frac{1}{N} \sum_{l=1}^{N} a_l = \lim_{l \rightarrow \infty} a_l,
\label{eq:Cesaro_summation}
\ee
provided the latter limit exists. Replacing $\tensor*[_{NS}]{\Braket{0,h_0|\sigma^x_n(t) \sigma^x_{n+l}(t)|0,h_0}}{_{NS}}$ with $\rho^{xx}(l,t)$ into Eq.~\eqref{eq:squared_m_intermediate} and then using Eq.~\eqref{eq:Cesaro_summation}, one has
\be
m^2_{\mathrm{ness}}=
\frac{1}{\widehat{q}(0)}\int_{0}^{\infty}\mbox{d}\tau \, q(\tau) \Braket{\sigma^x (\tau)}^2, 
\label{eq:squared_m_final}
\ee
where
\be
\Braket{\sigma^x (t)}^2 = \lim_{N \rightarrow \infty} [\mbox{Re}\left(\tensor*[_{NS}]{\Braket{0,h_0|\mathcal{\sigma}^x(t)|0,h_0}}{_R} \right)]^2
\label{eq:magnetization_odd_one_point}
\ee
according to Eq.~\eqref{eq:odd_operators_Z2_symmetry}, and we have used the \textit{clustering property} of correlation functions 
\be
\lim_{l \rightarrow \infty} \rho^{xx}(l,t) = \Braket{\sigma^x(t)}^2.
\label{eq:clustering_correlations}
\ee
Equation \eqref{eq:squared_m_final} coincides with Eq.~(9) of the main text and it is the final result that we have evaluated numerically. We report in the following all the expressions necessary for the evaluation of $m^2_{\mathrm{ness}}$ as in Fig.~2 of the main text. 

To compute $\Braket{\sigma^x}$ we follow the analysis of Refs.~\onlinecite{McCoy1,McCoy2,McCoy3,McCoy4}, where one considers the two point function in the large distance limit $l \rightarrow \infty$ according to Eq.~\eqref{eq:clustering_correlations}. The equal time two-point function of the longitudinal magnetization is non-trivial to compute because $\sigma^x$ is a non-local operator in terms of the Jordan-Wigner fermions because of Eq.~\eqref{eq:JW_transformation}
\begin{equation}
\sigma^x_n = \prod_{j=1}^{n-1} (i a^y_j a^x_j) a^x_j,
\label{eq:JW_sigma_x}
\end{equation}
where we have introduced the Majorana fermion operators $a^x_n$ and $a^y_n$ at site $n$, which are defined as
\begin{equation}
a_n^x = c_n+c_n^{\dagger}, \quad a_n^y=i(c_n^{\dagger}-c_n).
\label{eq:Majorana_operators}
\end{equation}
These operators obey the algebra 
\be
\{a_n^x, a_m^x \}=\{a_n^y, a_m^y \}=2\delta_{mn}, \quad \mbox{and} \quad \{a_n^x, a_m^y \}=0.
\label{eq:Majoran_operators_algebra}
\ee
Inserting Eq.~\eqref{eq:JW_sigma_x} with the Majorana fermions in Eq.~\eqref{eq:Majorana_operators} into Eq.~\eqref{eq:correlator_thermodynamic_limit}, one realizes from the anticommutation relation in Eq.~\eqref{eq:Majoran_operators_algebra} that only the string of Majorana operators between $i$ and $i+n$ matters 
\be
\rho^{xx}(l,t) = \lim_{N \rightarrow \infty} \tensor*[_{NS}]{\Braket{0,h_0|\prod_{n=1}^l (-i a_n^y(t) a_{n+1}^x(t))|0,h_0}}{_{NS}}.
\ee
Crucially in the expectation value of the previous expression the same state $\ket{0,h_0}_{NS}$ appears in the bra and in the ket. The Wick theorem can be therefore applied, which allows to reduce the calculation of the $n$-point correlation function of the operators $a_j^{x,y}$ in terms of the calculation of the two-point correlator of the same operators. This approach has been first developed in Refs.~\onlinecite{McCoy1,McCoy2,McCoy3,McCoy4}, and later applied in Refs.~\onlinecite{calabrese2011quantum,calabrese2012quantum1,calabrese2012quantum2} for quantum quenches of the transverse field, whose analysis we here follow. The resulting expression for the correlation function $\rho^{xx}(l,t)$ is
\be
\rho^{xx}(l,t)=\mbox{Pf}(\Gamma),
\label{eq:Pfaffian}
\ee
where $\mbox{Pf}(\Gamma)$ is the Pfaffian of the $2l \times 2l$ antisymmetrix matrix $\Gamma$. The latter is a block Toeplitz matrix (entries depend on the difference between the row and column index)
\begin{equation}
\Gamma = \begin{pmatrix}
\Gamma_0 & \Gamma_{-1} & \dots & \Gamma_{1-l} \\
\Gamma_1 & \Gamma_{0} &  \dots &  \dots \\
\dots & \dots & \dots & \dots \\
\Gamma_{l-1} & \dots & \dots & \Gamma_0
\label{eq:Gamma_matrix}
\end{pmatrix}        
,
\end{equation}
with
\begin{equation}
\Gamma_n =
\begin{pmatrix}
-f_n & g_n \\
-g_{-n} & f_n
\end{pmatrix}
.
\label{eq:Gamma_matrix_bloch}
\end{equation}
and the functions $f_n$ and $g_n$ are directly related to the two-point correlator of the Majorana operators 
\begin{equation}
f_n= -i\delta_{n0}+i \, \, {}_{NS}\Braket{0,h_0|a^x_j a^x_{j+n}|0,h_0}_{NS}, \quad \mbox{and} \quad g_n=i \, \, {}_{NS}\braket{0,h_0|a^x_{j} a^y_{j+n-1}|0,h_0}_{NS}.
\label{eq:propagator}
\end{equation}
The expression for $f_n$ and $g_n$ is particularly simple in Fourier space
\begin{equation}
f_n = \int_{-\pi}^{\pi}\frac{\mbox{d}k}{2 \pi} e^{ink} f(k), \quad g_n = \int_{-\pi}^{\pi}\frac{\mbox{d}k}{2 \pi} e^{ink} g(k),     
\end{equation}
with the Fourier transform $f(k)$ and $g(k)$ given by
\begin{equation}
f(k)= i \, \mbox{sin} \, \Delta_k \, \mbox{sin}(2 \varepsilon_h(k)t), \quad \mbox{and} \quad
g(k) = -e^{i \theta_h(k)-ik}[\mbox{cos} \, \Delta_k -i \mbox{sin} \, \Delta_k \, \mbox{cos}(2 \varepsilon_h(k)t)].  
\label{eq:T_H_functions}
\end{equation}
We report here for completeness the functions defined in Eq.~\eqref{eq:T_H_functions}
\begin{subequations}
\begin{align}
\mbox{cos} \, \Delta_k &= \frac{4J^2(h h_0 -(h+h_0)\mbox{cos}k +1)}{\varepsilon_h(k)\varepsilon_{h_0}(k)},  \\
\mbox{sin} \, \Delta_k   &=  \frac{4J^2(h-h_0) \mbox{sin}k}{\varepsilon_h(k)\varepsilon_{h_0}(k)},  \\
e^{i \theta_h(k)} &= \frac{2J(h-e^{ik})}{\varepsilon_h(k)},
\end{align}
\label{eq:f_k_and_g_k_functions}%
\end{subequations}
with the single quasi-particle dispertion relation $\varepsilon_h(k)$ given in Eq.~\eqref{eq:TFIC_spectrum}.
To perform this numerical analysis we have further exploited the fact that the calculation of the Pfaffian of the $2l\times 2l$ matrix $\Gamma$ can be reduced to the calculation of the determinant of an $n \times n$ matrix having the same spectrum as $\Gamma$, as first shown in Refs.~\onlinecite{calabrese2011quantum,calabrese2012quantum1,calabrese2012quantum2}. We report here the main steps underlying this reduction procedure since we have used it in our numerical analysis. The matrix $\Gamma$ is real and antisymmetric and therefore it has a spectrum composed of complex-conjugate pairs of eigenvalues $\pm i \lambda_m$, with $m=1,2 \dots n$. We define the $l \times l$ matrices $W_{\pm}$ as 
\be
W_{\pm} = T \pm i H, \quad \mbox{with} \quad \left\{
  \begin{array}{lr}
    T_{jn} = f_{j-n}, \quad \, \, \, \, \, \, \, \, \, \, \, \, \, \, j,n=1,2 \dots l \\
    H_{jn} = g_{j+n-l-1} \quad \, \, j,n=1,2 \dots l.
   \end{array}    
   \right.
   \label{eq:T_H_matrices}
\ee 
$T$ is a Toeplitz matrix and it is antisymmetric since $f_{-n}=-f_n$, while $H$ is an Hankel matrix (the entries depend on the sum of the row and column indexes) and it is therefore symmetric. It follows that $W_{\pm}^{\dagger}=-W_{\pm}$ are anti-Hermitean matrices. The spectrum of $W_{\pm}$ is therefore purely imaginary. In particular, the spectrum of $W_{+}$ is the complex conjugate of the spectrum of $W_{-}$. Given an eigenvector $\Vec{w}^{(m)}_j$ ($j=1,2 \dots l$) of $W_{+}$ with eigenvalue $i\lambda_m$, we have that the $2l$ dimensional vector $\Vec{v}^{(m)}$
\begin{equation}
\Vec{v}^{(m)}= \left\{
  \begin{array}{lr}
    \Vec{v}^{(m)}_{2j-1} = w_j^{(m)}, \\
    \Vec{v}^{(m)}_{2j} = -i w_{l+1-j}^{(m)},
   \end{array}    
   \right.    
\end{equation}
is an eigenvector of $\Gamma$ with eigenvalue $-i\lambda_m$. From the previous equation, by exploiting the relation $[\mbox{Pf}(\Gamma)]^2=\mbox{det}(\Gamma)$, one eventually has 
\be
\rho^{xx}(l,t) = (-1)^{l(l-1)/2} |\mbox{det}(W_{+})|.
\label{eq:correlator_numerics_expression}
\ee
Equations \eqref{eq:T_H_matrices} and \eqref{eq:correlator_numerics_expression} have been used in the manuscript to compute $\rho^{xx}(l,t)$ as a function of time $t$. Upon evaluating $\rho^{xx}(l,t)$ for large values of $l$ (we have used $l=200$ in our numerical analysis) one can extract the order parameter $\braket{\sigma^x(t)}^2$ as a function of time from the clustering decomposition formula in Eq.~\eqref{eq:clustering_correlations}. The resulting expression for $\braket{\sigma^x(t)}^2$ can be eventually inserted in Eq.~\eqref{eq:squared_m_final} to get $m^2_{\mathrm{ness}}$. The result is shown in Fig.~2 of the main text. We emphasize that the determinant representation of the correlation function in Eq.~\eqref{eq:correlator_numerics_expression} works for arbitrary values of both the pre-quench $h_0$ and the post-quench $h$ field. It can be therefore used both to study quenches within the ferromagnetic/paramagnetic phases and to study quenches across the two phases. For the case of interest in this manuscript, the pre-quench field is always fixed to zero, $h_0=0$ (cf., Eq.~\eqref{eq:reset_states_Z_2_symmetry}), while $h$ can assume arbitrary values. With our conditional reset protocol starting from the reset states in Eq.~\eqref{eq:reset_states_Z_2_symmetry}, the unitary time evolution is therefore described by either quantum quenches within the ferromagnetic phase ($h_0=0$ and $h<1$), or by quenches from the ferromagnetic to the paramagnetic phase ($h_0=0$ and $h>1$).

\subsection{Asymptotics of the magnetization density squared for large $J t_{\mathrm{max}}$ and low $\gamma/J$}
\label{sec:TFIC_asympt}

In this Subsection we eventually discuss the asymptotic behavior of the $m^2_{\mathrm{ness}}$ for large $J t_{\mathrm{max}}$ and low $\gamma/J$. This limit is physically interesting because in this regime the crossover displayed by $m^2_{\mathrm{ness}}$ as a function of $h$ becomes very sharp and it closely resembles the equilibrium quantum phase transition behavior in Eq.~\eqref{eq:GS_magnetization}. The result of this analysis is reported in Fig.~2(d) of the main text. We provide here all the necessary expressions for the underlying numerical evaluation.

In the limit of low $\gamma/J$ and large $J t_{\mathrm{max}}$, on average, a large time elapses between one reset and the subsequent one and therefore the integral in Eq.~\eqref{eq:squared_m_final} is dominated by the long time behavior of $\Braket{\sigma^x}^2(t)$. This is fundamental from the analytical point of view because the long time behavior of $\Braket{\sigma^x}(t)$ after a quantum quench displays universal features.  
In particular, for a quench within the ferromagnetic phase, $h_0, h<1$, it has been analytically proved in Refs.~\onlinecite{calabrese2011quantum,calabrese2012quantum1,calabrese2012quantum2} that the order parameter $\Braket{\sigma^x}(t)$ at long times $t\gg 1$ decays exponentially in time as 
\begin{equation}
\Braket{\sigma^x}(t) = \left(C_{FF}\right)^{1/2}\mbox{exp}\left(-t \Lambda(h,h_0)\right), \quad \mbox{as} \, \, t \to \infty,    
\label{eq:orderP_FF_1} 
\end{equation}
with
\begin{equation}
C_{FF}(h,h_0) = \frac{1-h h_0+\sqrt{(1-h^2)(1-h_0^2)}}{2\sqrt{1-hh_0}(1-h_0^2)^{1/4}},
\label{eq:amplitude_FF}
\end{equation}
and
\begin{equation}
\Lambda(h,h_0)= -\int_{0}^{\pi} \frac{\mbox{d}k}{\pi} \varepsilon_h'(k) \mbox{ln}|\cos \Delta_k|>0,
\label{eq:rate_FF}
\end{equation}
where $\varepsilon_h'(k) = \frac{d}{dk} \varepsilon_h(k)$. In our case $h_0=0$ and therefore we can restrict ourselves to the consideration of $\Lambda(h,0)$, which is given upon performing the integral in Eq.~\eqref{eq:rate_FF}
\begin{equation}
\Lambda(h,0)= \frac{1}{\pi}\left[4Jh-4J\sqrt{1-h^2}\left(\frac{\pi}{2}-2\arctan\left(\sqrt{\frac{1-h}{1+h}}\right)  \right)\right], \, \, \, \mbox{for} \, \, \, h<1.
\label{eq:rate_FF_h0}    
\end{equation}
Inserting Eqs.~\eqref{eq:orderP_FF_1}-\eqref{eq:rate_FF_h0} into Eq.~\eqref{eq:squared_m_final}, with the survival probability in Eq.~\eqref{eq:survival_chop_exp}, one has 
\begin{equation}
m^2_{\mathrm{ness}}=\frac{\gamma C_{FF}(h,0) e^{-\gamma t_{\mathrm{max}}}}{1-e^{-\gamma t_{\mathrm{max}}}-\gamma t_{\mathrm{max}}e^{-\gamma t_{\mathrm{max}}}}\left[\frac{e^{\gamma t_{\mathrm{max}}}-1}{\gamma +2\Lambda(h,0)}+\frac{\gamma(e^{-2 t_{\mathrm{max}} \Lambda(h,0)}-1)}{2 \Lambda(h,0)(\gamma+2\Lambda(h,0))} \right] \,\,\, \mbox{for} \,\, \, h<1.
\label{eq:analytic_m_2_FF}
\end{equation}
In the case $h>1$ (for quenches across the quantum critical point) the formula in Eq.~\eqref{eq:orderP_FF_1} does not apply and no analytical formula has been so far proved. In Ref.~\onlinecite{calabrese2012quantum1} the following conjecture has been put forward for long times $t \gg 1$
\begin{equation}
\Braket{\sigma^x}(t)=\left(C_{FP} \right)^{1/2} \left[1+\cos(2 \varepsilon_h(k_0)t+\alpha)+\dots \right]^{1/2} \mbox{exp}\left(-t \Lambda(h,h_0) \right),\quad \mbox{as} \, \, t \to \infty,
\label{eq:OrderP_FP_1}
\end{equation}
where the $\dots$ denotes subleading corrections. The constant $\alpha(h,h_0)$ in Eq.~\eqref{eq:OrderP_FP_1} is an undetermined phase shift, which has to be fixed by keeping it as a fitting parameter. The momentum $k_0$ is determined such that $\cos \Delta_{k_0}=0$. The amplitude $C_{FP}$ is 
\begin{equation}
C_{FP} = \left[\frac{h \sqrt{1-h_0^2}}{h+h_0}\right]^{1/2}, 
\label{eq:constant_FP_h0}
\end{equation}
Our aim is to use Eq.~\eqref{eq:OrderP_FP_1} to get an estimate of the asymptotic value attained by $m^2_{\mathrm{ness}}$ at large values of $h$ deep in the paramagnetic phase. By plugging Eq.~\eqref{eq:OrderP_FP_1} into Eq.~\eqref{eq:squared_m_final} one obtains at leading order as $h \rightarrow \infty$
\begin{equation}
m^2_{\mathrm{ness}}=\frac{\gamma e^{-\gamma t_{\mathrm{max}}}}{1-e^{-\gamma t_{\mathrm{max}}}-\gamma t_{\mathrm{max}}e^{-\gamma t_{\mathrm{max}}}}\left[\frac{e^{\gamma t_{\mathrm{max}}}-e^{-2 t_{\mathrm{max}} \Lambda(h,0)}}{\gamma +2\Lambda(h,0)}+ \frac{e^{-2 t_{\mathrm{max}} \Lambda(h,0)}-1}{2\Lambda(h,0)} \right], \,\,\, \mbox{as} \,\, \, h \rightarrow \infty,
\label{eq:analytic_m_2_FP}    
\end{equation}
with 
\begin{equation}
\Lambda(h,0)= \frac{4 J}{\pi}, \quad \mbox{for} \quad h>1,
\label{eq:rate_paramagnetic_h0}
\end{equation}
and $C_{FP}(h,0)=1$ from Eq.~\eqref{eq:constant_FP_h0}.
Equations \eqref{eq:analytic_m_2_FF} and \eqref{eq:analytic_m_2_FP} are reported in Fig.~2(d) of the main text as dashed lines for $h<1$ and $h>1$, respectively. A couple of important observations are now in order. First, $m^2_{\mathrm{ness}}$ is always different from zero, also for large values of $h$, for any finite value of $\gamma$ and/or $t_{\mathrm{max}}$. This is a direct consequence of the fact that $\rho_{\mathrm{ness}}$ is different from the diagonal ensemble, as commented after Eq.~\eqref{eq:Diagonal_ensemble_final_operator}. The diagonal ensemble is in the case here of interest equivalent to the generalized Gibbs ensemble, for local operators in the thermodynamic limit, since the TFIC is an integrable model \cite{calabrese2012quantum1}. The generalized Gibbs ensemble, in particular, predicts a vanishing expectation value for the magnetization density in the stationary state attained at long times after a quench. The long-range order of the pre-quench ground state is therefore entirely wiped out by the purely unitary quench time evolution. The non-zero value of $m^2_{\mathrm{ness}}$ that we computed in Fig.~2 is, therefore, an intrinsic property of the reset protocol, which shows that superimposing a controlled source of dissipation on top of the unitary time evolution eventually allows for the storage of part of the long-range order of the states $\ket{1}$ and $\ket{2}$ in the steady state. This information storage is not lost even for large values of $h$, deep in the disordered phase, where $m^2_{\mathrm{ness}}$ attains a finite limiting value in good agreement with the prediction of Eq.~\eqref{eq:analytic_m_2_FP}.
Second, we note that deviations from the asymptotics behavior predicted by \eqref{eq:analytic_m_2_FF} and \eqref{eq:analytic_m_2_FP} are visible only for $h \gtrsim 0.9$. In the latter case the discrepancy is caused by the fact that $\Braket{\sigma^x}(t)$ for large quenches, $h \gg h_0=0$, decays extremely rapidly as a function of time $t$ and it is non-zero only for short times. The time integral in Eq.~\eqref{eq:squared_m_final} is therefore not anymore dominated by the long time behavior, where equations \eqref{eq:orderP_FF_1} and \eqref{eq:OrderP_FP_1} apply. In the case $h>1$ one has a relative error of $10-15\%$ between the asymptotic value in Eq.~\eqref{eq:analytic_m_2_FP} at large $h$ and the numerically exact asymptotic value. We again remark that Eq.~\eqref{eq:OrderP_FP_1} is a conjecture, not rigorously proven, and therefore Eq.~\eqref{eq:analytic_m_2_FP} can be taken as an estimate of the residual (minimum) magnetization attained by the system for large values of the post-quench field $h$.

\bibliography{biblioPRB}